\documentclass[pra, twocolumn, aps, superscriptaddress,longbibliography,showpacs,amsmath,amssymb,floatfix]{revtex4-2} 

\usepackage[linesnumbered,ruled,vlined]{algorithm2e}
\usepackage{algorithmic}

\usepackage{graphics}
\usepackage{amssymb}
\usepackage{amsmath}
\usepackage{caption}
\usepackage{subcaption}
\usepackage{epsfig}
\usepackage{color}
\usepackage{tabu}
\usepackage{mathtools}
\usepackage[colorlinks,linkcolor=blue,anchorcolor=blue,citecolor=blue,urlcolor=blue]{hyperref}
\usepackage{physics}
\usepackage{float}
\usepackage{diagbox}
\usepackage{booktabs} 
\usepackage{makecell}

\usepackage{bm}

\usepackage{tikz}
\usetikzlibrary{quantikz}
\usetikzlibrary{calc,positioning}
\newcommand{\casql}{Laboratory of Quantum Information, University of Science and Technology of China, Hefei 230026, China}

\newcommand{\aihf}{Institute of Artificial Intelligence, Hefei Comprehensive National Science Center, Hefei, Anhui, 230088, China}
\newcommand{\anhuiqn}{Anhui Province Key Laboratory of Quantum Network,
University of Science and Technology of China, Hefei 230026, China}
\newcommand{\origin}{Origin Quantum Computing, Hefei, Anhui, 230088, China}
\newcommand{\IAT}{Institute of the Advanced Technology, University of Science and Technology of China, Hefei, Anhui, 230088, China}

\newcommand{\GateSWAP}{\textsf{SWAP}\xspace}
\newcommand{\GateCSWAP}{\textsf{CSWAP}\xspace}

\newcommand{\GateIntSWAP}{\textsf{Internal-SWAP}}
\newcommand{\GateRouting}{\textsf{Routing}}

\newcommand{\GateZ}{\textsf{Z}}
\newcommand{\GateY}{\textsf{Y}}
\newcommand{\GateX}{\textsf{X}}
\newcommand{\GateI}{\textsf{I}}
\newcommand{\GateH}{\textsf{H}}

\newcommand{\GateCZ}{\textsf{CZ}}

\newtheorem{dfn}{Definition}

\iffalse

\else

\fi

\iffalse

\else

\fi

\begin{document}
\title{Refined Criteria for QRAM Error Suppression via \\ Efficient Large-Scale QRAM Simulator}

\author{Yun-Jie Wang}
\affiliation{\IAT}
\affiliation{\aihf}
\affiliation{\casql}
\affiliation{\anhuiqn}

\author{Tai-Ping Sun}
\affiliation{\casql}
\affiliation{\anhuiqn}

\author{Xi-Ning Zhuang}
\affiliation{\casql}
\affiliation{\anhuiqn}

\author{Xiao-Fan Xu}
\affiliation{\casql}
\affiliation{\anhuiqn}

\author{Huan-Yu Liu}
\affiliation{\aihf}

\author{Cheng Xue}
\affiliation{\aihf}

\author{Yu-Chun Wu}
\affiliation{\IAT}
\affiliation{\aihf}
\affiliation{\casql}

\affiliation{\anhuiqn}

\author{Zhao-Yun Chen}
\affiliation{\aihf}

\author{Guo-Ping Guo}
\affiliation{\IAT}
\affiliation{\aihf}
\affiliation{\casql}
\affiliation{\anhuiqn}
\affiliation{\origin}

\begin{abstract}
Quantum random access memory (QRAM) is a critical primitive for quantum algorithms that require data lookup in superposition, but its lack of fault tolerance poses a major obstacle to practical deployment. Error filtration (EF) has been proposed as a hardware-efficient alternative to error correction, capable of suppressing incoherent noise without encoding overhead. However, its performance in realistic QRAM systems with moderate fidelity has remained unclear, as existing analyses rely on asymptotic approximations and numerical simulations have been limited to small sizes.
We address this gap using a new simulator for bucket-brigade (BB) QRAM that combines sparse state encoding with a noise-aware pruning algorithm. This framework provides full quantum state access and scales efficiently, enabling us to probe EF performance in size and noise regimes far beyond previous studies. Our simulations reveal suppression anomalies at high noise levels or large address sizes, where post-selection probability fundamentally constrains EF scaling. Incorporating this effect, we refine EF theory into near-deterministic criteria linking base infidelity to achievable suppression, thereby delineating the regime in which EF yields progressive improvement.
Beyond refining EF, we quantitatively characterize the runtime and memory costs of our noisy BB QRAM simulator, achieving simulations of systems with 20 layers using less than 1 GB of memory.
This efficiency is what enables us to probe parameter regimes beyond previous work and to establish the simulator as a practical, ``fine-print'' analysis tool for assessing QRAM as a quantum resource.
\end{abstract}

\maketitle

\section{Introduction}

Computers are fundamentally data-processing machines, and their performance has always been shaped by how efficiently they can access memory.
In classical computing, the development of random access memory (RAM) enabled scalable architectures by allowing fast, indexed retrieval of data, becoming a cornerstone of modern computing systems~\cite{jaeger2011microelectronic}. As quantum computing advances~\cite{Grover_1996, Shor_1999, Preskill_NISQ_2018, Megaquop_Preskill_2025}, it is natural to ask whether similar memory-access structures can be developed to support quantum data processing. 
Many quantum algorithms, ranging from linear algebra~\cite{HHL_2009, Pre_QLSA_2013, QSLP_Childs_2017} and chemistry simulations~\cite{ChemDynamics_Alan_2008, Truncated_TS_2015, Babbush_Spectra_2018, QChem_in_QC_2019} to quantum machine learning~\cite{QSVM_PRL_2014, Lloyd2014, Biamonte_QML_2017}, rely on the ability to perform data lookups coherently.
A quantum random access memory (QRAM) would provide exactly this capability, serving as the quantum analogue of classical RAM and as a critical resource for future large-scale quantum algorithms.



Among the candidate architectures~\cite{Asaka_2021, Two_Level_Walker_QRAM, Park2019, Spin_Photon_2021, Dirk_QRAMNetworkSimulation_2023, Mukhopadhyay2025, deriso2025resourceefficientquantumwalkerquantumram}, the bucket-brigade (BB) QRAM stands out for its appealing theoretical properties: logarithmic query depth with respect to memory size~\cite{Seth_BBQRAM_2008, Seth_BBQRAMArch_2008}, and infidelity that scales only polylogarithmically with memory size as well~\cite{Connor_QRAMNoise_2021, Connor_QRAMThesis_2021, mehta2024analysissuppressionerrorsquantum}. 
These features have made it the focus of extensive theoretical work and recent experimental proposals~\cite{YongShan_QRAMSystem_2023, Weiss_3DCavityQRAM_2024, YunJie_2025, transmon_phonon_qram_2025, zhang2025demonstratingcoherentquantumrouters, shen2025experimentalrealizationbucketbrigadequantum}. The BB architecture, together with its experimental instantiations, also represents an effort to address the ``read the fine print'' critique~\cite{Scott_FinePrint_2015}. Yet, despite this promise, the feasibility of QRAM in realistic settings remains unresolved. Concerns persist about its fault tolerance, resource requirements, and practical utility in near-term devices~\cite{liu2023quantummemorymissingpiece, jaques_qramsurveycritique_2023}.

A central obstacle to realizing QRAM is fault tolerance.
With the exception of a few non-additive cases, no binary quantum error-correcting code can transversally implement a universal reversible gate set~\cite{Yaoyun_transversal_2018}, which implies extreme overheads for fully fault-tolerant execution of BB QRAM, whose basic operations are the reversible gates $\{\GateSWAP, \GateCSWAP\}$.
Surface-code resource estimates suggest that even a modest 30-layer BB QRAM would require $10^{15}$ physical qubits for millisecond-scale queries, or millions of qubits with query times on the order of years~\cite{Olivia_resource_qram}.
These daunting figures have fueled skepticism about near-term practicality.

Several mitigation strategies have been proposed: treating BB QRAM as a noisy black box within fault-tolerant workflows~\cite{dalzell2025distillationteleportationprotocolfaulttolerantqram}, improving its internal design to reduce error accumulation~\cite{YongShan_QRAMSystem_2023, mehta2024analysissuppressionerrorsquantum}, or exploiting its favorable noise scaling without full error correction~\cite{Gideon_EFiltration_2023}.
Among the latter, error filtration (EF)~\cite{Gideon_EFiltration_2023} has emerged as a particularly promising approach: a hardware-efficient, gate-based method for suppressing incoherent noise through controlled repetition and post-selection.
Unlike full fault tolerance, EF requires no encoding overhead, making it especially attractive for near-term QRAM implementations where the output state must be preserved.

Yet all of these approaches share a common prerequisite: classical simulators capable of modeling noisy BB QRAM at scale with full quantum state access—playing a role analogous to Stim~\cite{Gidney2021stimfaststabilizer} for stabilizer circuits.
Conventional statevector simulators scale exponentially with qubit number, and the binary tree structure of BB QRAM compounds this cost, leading to memory demands as large as $\exp(2^n)$ for $n$ address qubits in naive implementations.
Existing BB QRAM simulators~\cite{Connor_QRAMNoise_2021, Connor_QRAMThesis_2021, YongShan_QRAMSystem_2023} have reached up to $n=12$, but their scope has been limited: most focus only on fidelity, without systematic benchmarking of runtime or memory, and without integration into higher-level quantum algorithms. As a result, these approaches evaluate QRAM in isolation, rather than assessing its role as a usable quantum resource.

In this work, we close this gap by developing a BB QRAM simulator that combines a sparse state representation with a noise-aware pruning algorithm, enabling large-scale noisy simulations with full quantum-state access. This capability makes BB QRAM usable as a concrete, traceable oracle within larger quantum algorithms. Our framework therefore goes beyond fidelity estimates alone: it benchmarks runtime and memory costs, and it supports application-level studies, providing a comprehensive and practical assessment of QRAM performance in algorithmic contexts.

A natural case study is error filtration (EF), which suppresses noise while preserving the full quantum state. This property, unlike quantum error mitigation~\cite{QEM_Review} that typically recovers only expectation values, makes EF particularly well aligned with the role of QRAM as an oracle within larger algorithms. EF has also been proposed as a hardware-efficient alternative to full quantum error correction, capable of reducing incoherent noise without encoding overhead~\cite{Gideon_EFiltration_2023}. Yet the range over which EF remains effective has not been clearly established, motivating the need for large-scale simulations. Using our simulator, we perform the first large-scale study of EF in QRAM, revealing suppression anomalies invisible to asymptotic EF theory. By incorporating post-selection probability into the analysis, we refine EF theory to yield nearly deterministic suppression predictions from base infidelity, providing practical criteria for when EF enhances QRAM performance.

The remainder of the paper is organized as follows.
Section~\ref{Sec: Preliminaries} reviews the BB QRAM architecture and the error filtration (EF) protocol.
Section~\ref{Sec: Method} describes our simulation framework, introducing the sparse encoding scheme and the branch pruning algorithm, and benchmarks the simulator, evaluating static costs (time and memory) and noise-induced overheads, including pruning performance.
Section~\ref{Sec: EF} analyzes EF in noisy QRAM, covering the emergence of anomalies, the role of post-selection probability, and refined scaling criteria.
Finally, Section~\ref{sec: Conclusions} summarizes our findings and discusses implications for future QRAM applications.

\section{Preliminaries \label{Sec: Preliminaries}}
\subsection{Bucket-Brigade QRAM \label{Subsec: Bucket-Brigade QRAM}}
Quantum random access memory (QRAM) enables coherent quantum queries in $\mathcal{O}(\log N)$ time steps, where $N = 2^n$ is the number of memory cells.  
To support multi-bit data values, the $(n,k)$-QRAM formalism~\cite{QRAM_Parallel_Chen_2023} extends the standard model by introducing a $k$-qubit data register alongside the $n$-qubit address register.

\begin{dfn}\label{def: QRAM transform}
An $(n,k)$-QRAM implements the unitary transformation
\begin{equation}
\sum_{i,j} \alpha_{i,j} \ket{i}_{A} \ket{j}_{D}
\ \longrightarrow \
\sum_{i,j} \alpha_{i,j} \ket{i}_{A} \ket{j \oplus d_i}_{D},
\end{equation}
where $\ket{i}_{A}$ and $\ket{j}_{D}$ are computational basis states of the address and data registers, respectively, $d_i$ is the $k$-bit classical data stored at address $i \in [0, 2^n-1]$, and $\alpha_{i,j}$ are arbitrary complex amplitudes.
\end{dfn}

A widely studied physical realization is the \emph{bucket-brigade} (BB) QRAM architecture, which uses an ancillary binary tree to route quantum data from the root to the appropriate memory cell.  
Each node of the tree contains two qudits: one storing address information and the other data.  
The address qudits may be encoded as either three-level systems ($\ket{W}$, $\ket{L}$, $\ket{R}$) or binary qubits ($\ket{0}$, $\ket{1}$), both of which exhibit an inherent degree of noise resilience~\cite{Connor_QRAMNoise_2021}.

\begin{figure}[ht]
    \centering
    \includegraphics[width=\columnwidth]{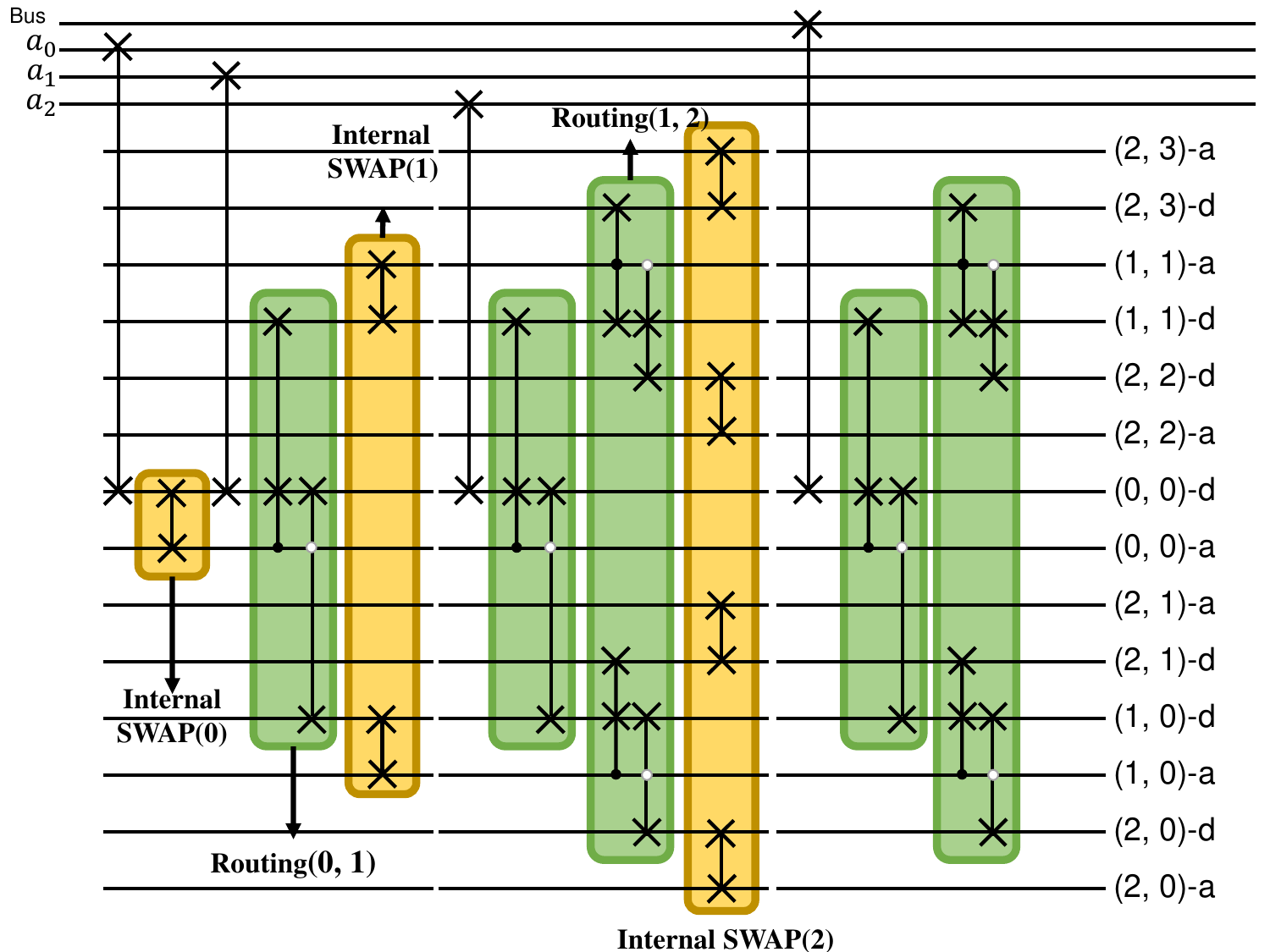}
    \caption{\textbf{BB QRAM fundamental operations.} \\
    Example for $n=3$ address qubits ($a_0$, $a_1$, $a_2$) and one data qubit ($d_0$).
    Layer-wise operations $\GateRouting$ (green) and $\GateIntSWAP$ (yellow) propagate address bits down the binary tree and route the data qubit to the correct memory cell.  
    Nodes are indexed as $(l, k)$-a/d, denoting the $k$-th node in layer $l$ for address/data registers.
    }
    \label{fig:QRAM Operations}
\end{figure}

The BB QRAM query protocol can be expressed in terms of two primitive operations:  
The swap gate $(\GateSWAP)$, which exchanges the states of two qubits, and the controlled-swap gate $(\GateCSWAP)$, which conditionally swaps two qubits based on a control qubit.  

Their \emph{layer-wise} analogues, $\GateIntSWAP$ and $\GateRouting$, operate in parallel across all nodes at a given tree depth, as illustrated in Fig.~\ref{fig:QRAM Operations}. In this figure, $\GateIntSWAP$ corresponds to ordinary $\GateSWAP$ gates in the qubit-based version, while in the qutrit-based implementation, these gates are conditioned on the address state of the parent node. The $\GateRouting$ operation, in contrast, retains the same structure across both encodings.
A full description of the BB QRAM query sequence is provided in Section~\ref{sec: Intro SI} of the Supplementary Information.

\subsection{Error Filtration and Suppression \label{Subsec: EF-Intro}}
Although the BB QRAM offers some intrinsic noise resilience~\cite{Connor_QRAMNoise_2021}, realistic implementations remain vulnerable to accumulated errors. 
Error filtration (EF)~\cite{Gideon_EFiltration_2023} provides a hardware-efficient protocol for suppressing such errors without requiring full quantum error correction.  
EF reduces infidelity by repeating a noisy operation in a coherently controlled fashion and post-selecting on an auxiliary register, thereby removing erroneous components at the cost of a reduced success probability.  
This makes EF particularly attractive for BB QRAM, where the output quantum state must be preserved after the query and standard fault-tolerance procedures are not yet practical.

\begin{figure}[ht]
    \centering
    \includegraphics[width=\columnwidth]{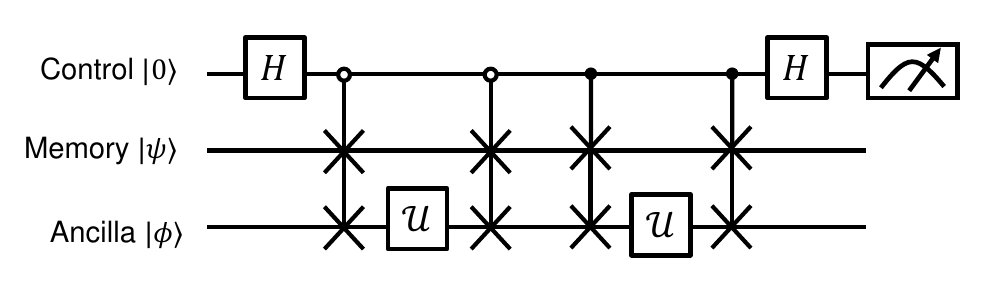}
    \caption{\textbf{Gate-based EF circuit for $T=1$.}\\  
    With $T=1$, the control register uses a single qubit and implements $2^1$ repetitions of the noisy operation $\mathcal{U}$.
    An auxiliary ancilla state $\ket{\phi}$ is used during the process and discarded at the end, while the control qubit is post-selected in the $\ket{0}$ state.  
    The resulting memory state retains only half of its original infidelity.
    }
    \label{fig:Error_Filtration_Circuit}
\end{figure}

The EF circuit for $T=1$ is shown in Fig.~\ref{fig:Error_Filtration_Circuit}, where $T$ denotes the EF level. 
For $T=1$, a single-qubit control register prepared in $\ket{+}$ coherently selects between two applications of the noisy operation $U$ acting on a \emph{memory} register ($\ket{\psi}$) and an \emph{ancilla} register ($\ket{\phi}$). 
After both controlled operations, the control is measured in the computational basis and post-selected on $\ket{0}$. 
The structure naturally generalizes: at level $T$, a $T$-qubit control register coordinates $2^T$ calls to $U$, and post-selection is performed on the all-zero state, ideally suppressing the infidelity by a factor of $2^T$.

\begin{figure*}[ht]
    \centering
    \includegraphics[width=\textwidth]{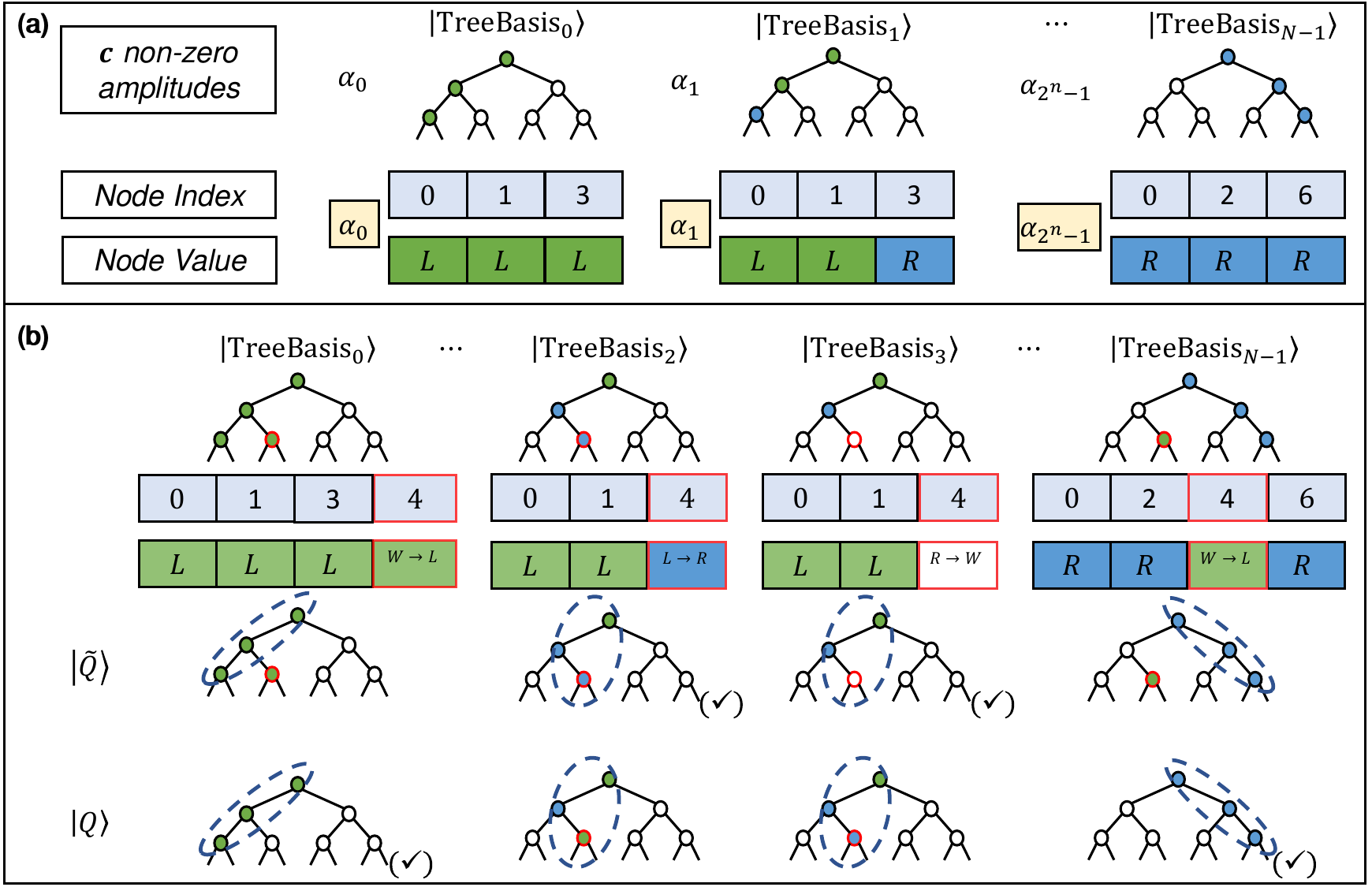}
    \caption{
    \textbf{Sparse representation and infidelity scaling of address-conditioned QRAM branches.} \\ 
    \textbf{(a)} Sparse branch encoding of BB QRAM states. Each address \( |a\rangle \) determines a unique routing path through the binary tree, represented here as a tree basis state \( |\text{TreeBasis}_c\rangle \) with amplitude \(\alpha_c\). 
    For each branch, only the active routing nodes along the path from root to leaf are stored. 
    The node index array (light blue boxes) records the fixed positions of active nodes, while the node value array (green/blue boxes) stores their values (e.g., L, R, W for qutrits). 
    All other nodes are implicitly in their idle state and omitted from storage. 
    This sparse-map representation eliminates the need to store the full Hilbert space vector, reducing memory to \( \mathcal{O}(n) \) per branch and enabling efficient layer-wise updates.\\
    \textbf{(b)} Illustration of the pruning algorithm using the subtree containment criterion. 
    The red-circled node marks the error location \( e = (l, p) \) at depth \(l\) and position \(p\). 
    For each branch \(|\mathrm{TreeBasis}_i\rangle\), we determine whether its routing path includes this node or any node in its subtree. 
    If so, the branch is \emph{unreliable} (top row, \(|\tilde{Q}_i\rangle\)); otherwise, it is \emph{reliable} (bottom row, \(|Q\rangle\)) and shares the same noiseless tree state and correct data output. 
    In this example, the fault affects only branches 2 and 3, while branches 0, 1, and the rest are unaffected and pruned from the noisy simulation.
    }
    \label{fig:QRAM_Simulator_illustration}
\end{figure*}

While EF has been proposed as a theoretically universal method for noise suppression, prior validations have been restricted to small systems (e.g., address sizes $n \leq 3$). 
Whether EF can sustain its predicted suppression factor of $2^T$ in larger circuits remains an open question. 
Using our scalable QRAM simulator, we extend the analysis far beyond previous limits, directly probing EF performance under increasing QRAM size and noise strength. 
These large-scale simulations uncover \emph{suppression anomalies} that are invisible in asymptotic treatments, motivating a refined analysis in realistic EF implementations.

\section{Efficient Simulation of Noisy QRAM \label{Sec: Method}}
The architectural features of the bucket-brigade (BB) QRAM enable significant simplifications in its classical simulation. Although it requires $\mathcal{O}(N)$ qubits and operations, its query protocol is built entirely from the reversible gate set $\{\GateSWAP, \GateCSWAP\}$, which act only as permutations of computational basis states. As a result of linearity, the same efficiency extends to initial states that are superpositions of only polynomially many basis states. Each query address then follows a unique, deterministic routing path through the binary tree, and in the absence of noise, different addresses evolve independently. These properties allow the simulation to be reorganized around \emph{branches}—the computational-basis paths defined by individual addresses—so that simulation cost scales with the number of active branches rather than the full $2^n$ state space.

In this section, we introduce two complementary techniques that together make large-scale BB QRAM simulation feasible. The first is sparse encoding, which exploits the tree-structured routing of BB QRAM to store only the minimal information required to represent an active branch, and to schedule operations layer-wise rather than gate-wise. Notably, this encoding scheme is fully compatible with the sparse data structures developed in Ref.~\cite{chen2023scalableprogramimplementationsimulation}, allowing us to seamlessly integrate a general-purpose sparse simulator with our QRAM-specific framework. The second is branch pruning, which leverages the locality of noise propagation in the binary tree to skip simulation of branches that are provably unaffected by faults. Sparse encoding provides the data representation and scheduling primitives that eliminate the exponential blow-up in memory requirements, while pruning reduces the number of branches that must be simulated under noise. Together, they form the algorithmic backbone of our simulator, enabling us to scale to address sizes, branch counts, and noise levels far beyond the reach of conventional state-vector methods.

\subsection{Sparse Encoding}
Conventional simulators represent the full quantum state of the QRAM circuit, requiring resources that scale exponentially with the address size. In contrast, our framework reorganizes the simulation around \emph{branches}, where each branch corresponds to a distinct address path through the BB QRAM’s binary routing tree. This branch-wise perspective, enabled by the structural regularities of the BB QRAM architecture, allows the global state to be expressed as a sparse collection of address-conditioned subspaces, eliminating the need to store or evolve the entire state vector.

A generic QRAM state can be written as
\begin{equation}
    \sum_{a,d,t} \alpha_{a,d,t} \, |a\rangle\, |d\rangle\, |\text{TreeBasis}\rangle,
    \label{eqn: Initial QRAM state decomposition}
\end{equation}
where \( |a\rangle \) encodes the address register, \( |d\rangle \) the data register, and \( |\text{TreeBasis}\rangle \) the collective state of the QRAM routing tree.

In BB QRAM, each address \( |a\rangle \) determines a unique routing path, making the tree configuration conditionally independent across addresses in the absence of noise. With this observation, we can reorganize the state as
\begin{equation}
    \sum_a |a\rangle \sum_c \alpha_c \, |d\rangle\, |\text{TreeBasis}_c\rangle,
    \label{eqn: address-conditioned branches}
\end{equation}
where \( |\text{TreeBasis}_c\rangle \) denotes the internal-node configuration for a given address branch.

In practice, we store each $\ket{\text{TreeBasis}}$ in a sparse map representation,as shown in Fig.~\ref{fig:QRAM_Simulator_illustration}(a), for a given branch.
\begin{itemize}
    \item Keys: node indices along the active path from root to leaf.
    \item Values: node basis states (e.g., L, R, W for qutrits; 0/1 for qubits).
    \item Amplitude: a single complex coefficient $\alpha$ per branch.
\end{itemize}

Sparse encoding not only minimizes memory use but also simplifies how operations are stored and applied.
Instead of applying every gate instance individually, we adopt a layer-wise operation abstraction: gates of the same type acting on all nodes at a given depth are grouped into a single instruction for that layer.
Examples include the \( \GateRouting \) and the \( \GateIntSWAP \), as shown in Fig.~\ref{fig:QRAM Operations}. This abstraction significantly reduces the number of instructions that need to be stored and interpreted during simulation.

To organize execution, we employ a two-level scheduling strategy. At the first level, the BB QRAM protocol is divided into $\mathcal{O}(n)$ discrete time steps, which are fixed once the address and data sizes are specified. Routing proceeds ballistically as a deterministic sequence of controlled gates~\cite{jaques_qramsurveycritique_2023}. At the second level, operations within each time step are scheduled layer by layer: all gates acting at the same tree depth are batched together. For each layer, the affected address range is precomputed, enabling direct access to the relevant entries in the sparse map and eliminating per-gate control overhead.

This compact representation eliminates the need to store the full state vector, instead retaining only the nonzero amplitudes and the minimal metadata required to reconstruct the active path.  
While conventional state-vector simulations require $\exp(N)$ memory and rapidly become intractable as the QRAM size grows, our sparse encoding enables a classical simulation algorithm with both space and time complexity scaling as $\mathrm{poly}(N)$.  
A more detailed discussion of how this new simulator relates to the two standard paradigms of classical quantum-circuit simulation is provided in Section~\ref{Sec: Sparse SI} of Supplementary Information.
Importantly, sparse encoding also facilitates efficient noise handling, laying the foundation for the pruning algorithm described next.
\begin{figure*}[ht]
    \centering
    \includegraphics[width=\textwidth]{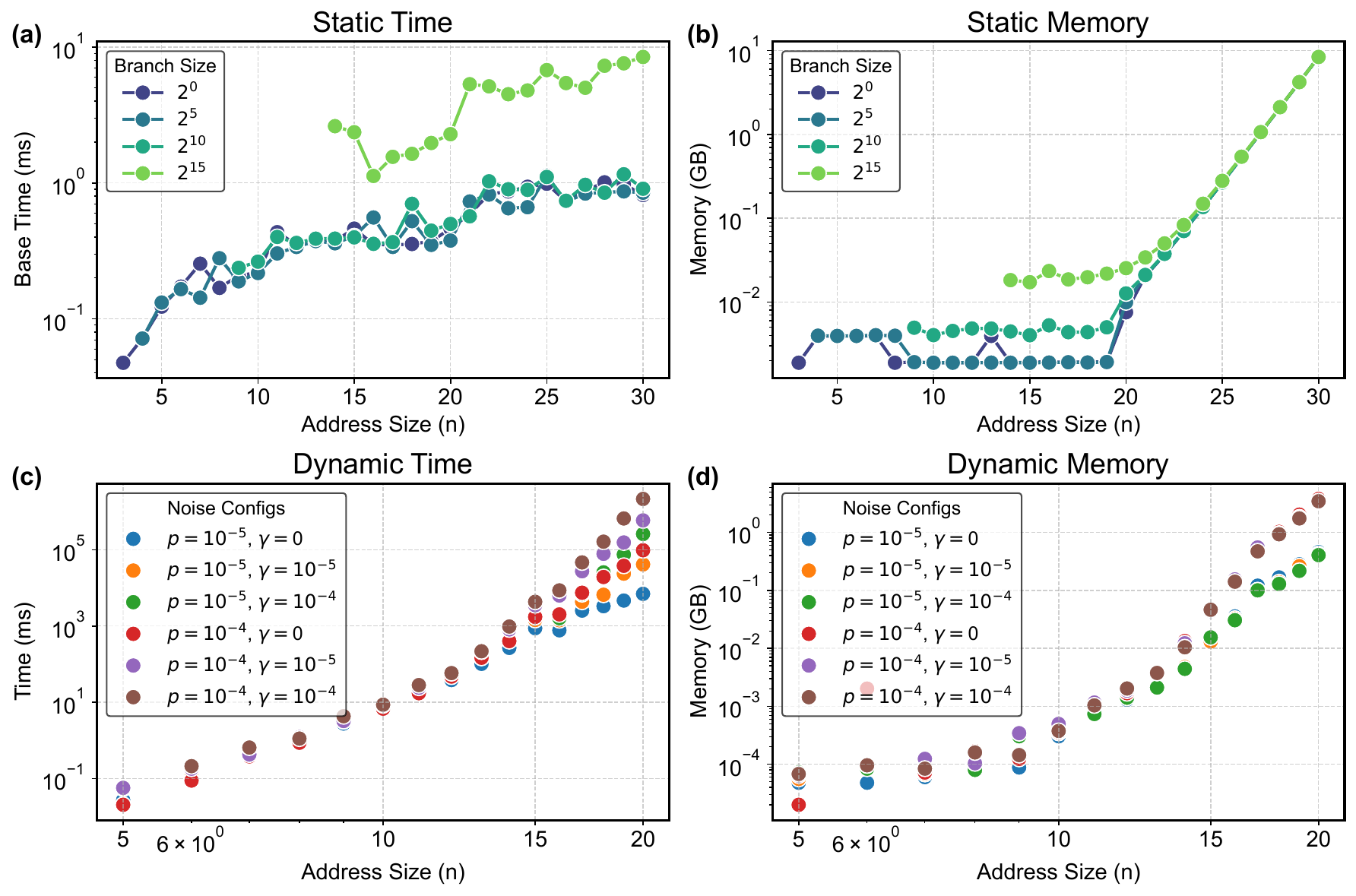}
    \caption{\textbf{Full-mode benchmark results for static and dynamic costs.} \\
    \textbf{(a)} Static runtime (noiseless, fixed branch size) as a function of address size \(n\), for branch sizes \(2^0\), \(2^5\), \(2^{10}\), and \(2^{15}\).  
    Runtimes remain flat with \(n\) for all branch sizes, confirming that noiseless cost depends only on branch size (number of nonzero amplitudes in the sparse encoding).  \\
    \textbf{(b)} Static memory usage in the noiseless case.  
    For small \(n\), memory is dominated by branch storage and remains flat; as \(n\) grows, the classical data term \(\mathcal{O}(2^n)\) dominates, causing convergence across branch sizes.\\  
    \textbf{(c)} Dynamic runtime---defined as the noisy runtime minus the static baseline---for various noise configurations \((p, \gamma)\).  
    Growth with \(n\) and \(p\) matches the theoretical scaling of infidelity, with clear separation into Regions~I–III (noise-free, transitional, and noise-dominated).\\  
    \textbf{(d)} Dynamic memory, defined analogously to (c), showing the same regime separation.  
    In both (c) and (d), the scaling with \(n\) and \(p\) reflects the predicted \(\mathcal{O}(n^2 p \, 2^n)\) dependence of the number of noisy branches.
    }
\label{fig:QRAM_FullMode_Benchmark}
\end{figure*}

\begin{figure*}[ht]
    \centering
    \includegraphics[width=\textwidth]{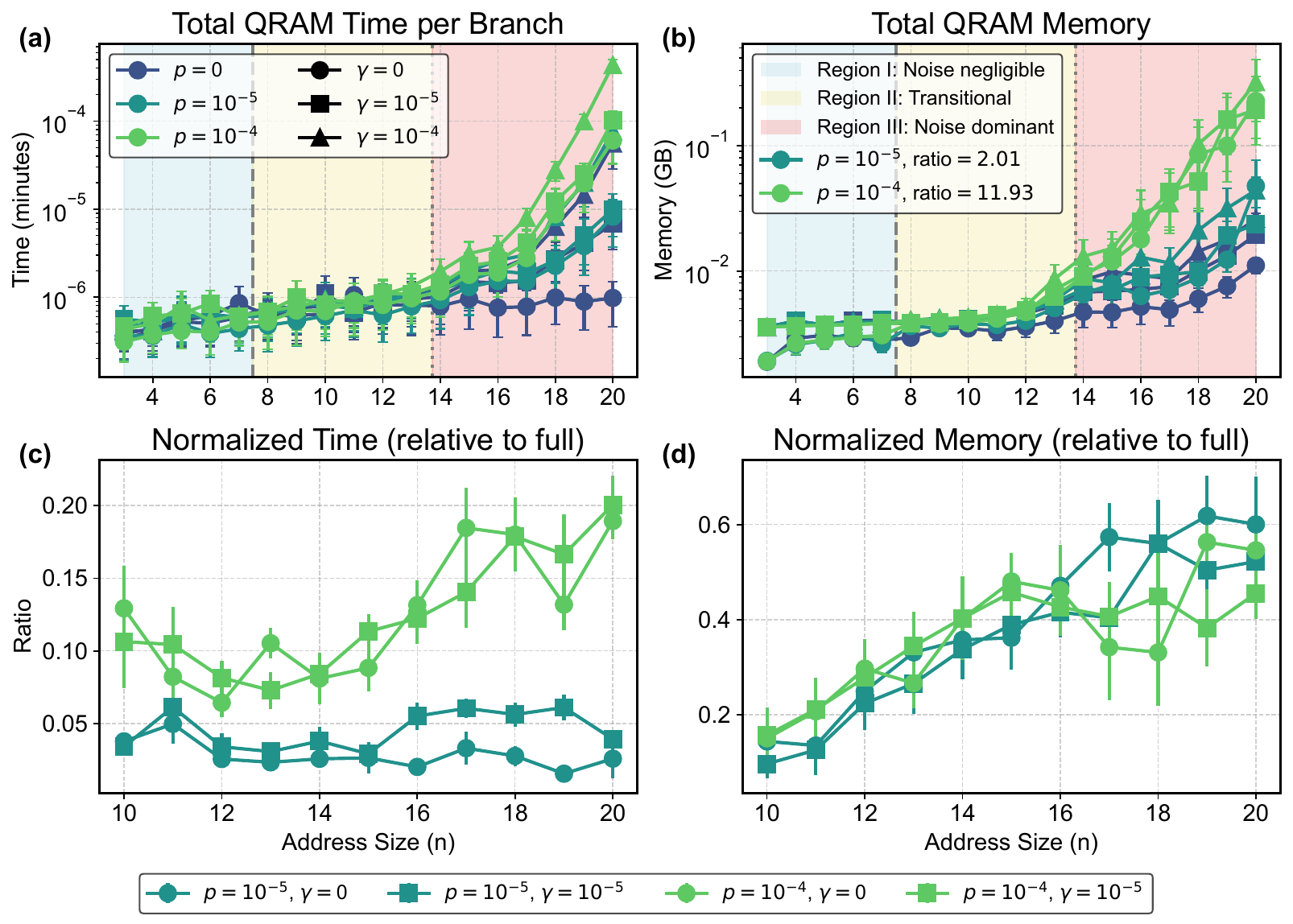}
    \caption{
    \textbf{Pruning-mode benchmark results and performance gains.}  \\
    \textbf{(a)} Total runtime per branch with pruning enabled, for various noise configurations \((p, \gamma)\). 
    Shaded backgrounds indicate the three scaling regimes from noise analysis: Region~I (noise-negligible, blue), Region~II (transitional, yellow), and Region~III (noise-dominated, red).  
    The same scaling trends as in full mode are preserved, but absolute runtimes are substantially reduced.  \\
    \textbf{(b)} Total memory usage in pruning mode, with noisy-to-noiseless memory ratios for \(p = 10^{-5}\) and \(p = 10^{-4}\) matching the linear scaling with \(p\) predicted by theory.  \\
    \textbf{(c)} Runtime ratio of pruning mode to full mode, showing reductions to as low as \(\sim 20\%\) of the full-mode cost in the noise-dominated regime at large \(n\).  \\
    \textbf{(d)} Memory ratio of pruning mode to full mode, with reductions to \(\sim 60\%\) in the same regime.  
    These results demonstrate that pruning achieves large and consistent cost savings without altering the underlying scaling behavior of noise-induced resource overhead.
    }
    \label{fig:QRAM_Benchmark_Total_Memory}
\end{figure*}

\subsection{Branch Pruning Algorithm}
In BB QRAM, a fault influences only those branches whose routing paths traverse the faulty node or its descendants. This subtree containment property means that the set of affected branches is precisely those passing through the faulty region, allowing all other branches to be skipped in simulation.

We formalize this by introducing the \emph{subtree containment criterion}:  
Formally, given a fault $e=(l,p)$ at depth $l$ and index $p\in[0,2^l-1]$, the set of affected addresses is
\begin{equation}
    i \in \mathcal{R}_T^{(l,p)} = \left[ L_T^{(l,p)},\ R_T^{(l,p)} \right],
\end{equation}
where
\begin{equation}
    L_T^{(l,p)} = 2^{n-l} \cdot p, \quad R_T^{(l,p)} = L_T^{(l,p)} + 2^{n-l} - 1.
\end{equation}
An address $i$ is affected if and only if $i \in \mathcal{R}_T^{(l,p)}$.

Branches outside $\mathcal{R}_T^{(l,p)}$ are reliable, meaning:
\begin{enumerate}
    \item Its data register contains the correct result.
    \item Its QRAM routing tree is in the same state \(|Q\rangle\) as all other reliable branches.
\end{enumerate}

For a randomly sampled noise configuration on the QRAM binary tree, we could determine the set \(S_{\mathrm{good}}\) of reliable addresses in advance and the final state of a single representative branch \(|Q_0\rangle\) will be:
\begin{equation}
    |i\rangle |d_i\rangle |Q_0\rangle, \quad i \in S_{\mathrm{good}}.
\end{equation}
Branches in \(S_{\mathrm{bad}}\) (the complement) are simulated individually with their own final tree states \(|\tilde{Q}_i\rangle\) corresponding to the check-marked configurations shown in Fig.~\ref{fig:QRAM_Simulator_illustration}(b).

For multiple faults at positions \(\{ e_1, e_2, \dots, e_k \}\), the set of unreliable branches is the intersection of the affected sets from each fault:
\begin{equation}
    \mathcal{R}_T = \bigcap_{j=1}^k \mathcal{R}_T^{(l_j, p_j)}.
\end{equation}
Only these branches are simulated under noise; all others are pruned (see Section~\ref{Sec: Noise Channels SI} of the Supplementary Information for details on the implementation of noise channels).

In total, the pruning algorithm enables substantial cost reduction while preserving physical accuracy.  
In the absence of noise, the cost scales as \(\mathcal{O}(d)\), while under noise it increases only by an additive term proportional to the number of affected branches, \(\mathcal{O}(p \, \mathrm{poly}(N))\).  
This scalability allows us to simulate BB QRAM circuits at sizes and noise levels far beyond the reach of full-state approaches.  
Crucially, the combination of \emph{sparse encoding} and \emph{noise pruning} makes large-scale BB QRAM simulation tractable.  

\subsection{Benchmark Results}
In this subsection, we present numerical benchmarks evaluating the performance of our QRAM simulator in both noiseless and noisy regimes. We first quantify core resource metrics, runtime and memory consumption, across varying address sizes, branch sizes, and noise strengths, establishing baseline scaling trends. We then evaluate the impact of the pruning algorithm, showing how it yields substantial reductions in computational cost. While these measurements are primarily a performance study, they also provide an empirical cross-check of our noise-propagation analysis: In a correctly implemented simulator, the additional runtime and memory overhead in the noisy case should arise almost entirely from the application of noise operators.

\subsubsection{Static Cost Baseline}
\paragraph{Time.} In noiseless simulations with fixed branch size, runtime is essentially independent of address size $n$ [Fig.~\ref{fig:QRAM_FullMode_Benchmark}(a)]. For branch sizes $2^0$–$2^{10}$ runtimes remain flat around $10^{-1}$–$10^0$ ms. For $2^{15}$, the runtime is proportionally larger but still insensitive to $n$. This confirms that noiseless cost depends only on branch size, not the total address space.

\paragraph{Memory.} Static memory has two parts: (i) branch storage (flat in $n$) and (ii) classical memory for the QRAM data table ($\mathcal{O}(2^n)$). At small $n$ the former dominates, while at large $n$ the latter overtakes, leading to convergence of curves across branch sizes [Fig.~\ref{fig:QRAM_FullMode_Benchmark}(b)]. The exponential data cost is unavoidable, but sparse encoding ensures that branch-related memory remains flat until the classical term dominates.

\subsubsection{Noise-Induced Overhead}
The dynamic cost, defined as $\Delta\text{Cost} = \text{Cost}_\text{noisy}-\text{Cost}_\text{noiseless}$, reflects noise-induced overhead. Figures~\ref{fig:QRAM_FullMode_Benchmark}(c,d) show runtime and memory overhead in full mode, while Figs.~\ref{fig:QRAM_Benchmark_Total_Memory}(a,b) show pruning mode. In all cases, scaling with $(n,p)$ follows theoretical predictions: the expected number of faulty nodes is $N\varepsilon$ with $N=\mathcal{O}(2^n)$, and each fault affects only its subtree, giving total affected branches $\mathcal{O}(N\varepsilon \, \mathrm{polylog}(N))$. The resulting infidelity therefore scales as $\mathcal{O}(\varepsilon \,\mathrm{polylog}(N))$, reflecting the resilience of BB QRAM.
We identify three regimes in~\ref{fig:QRAM_Benchmark_Total_Memory}(a, b):
\begin{enumerate} 
\item \textbf{Region I: Noise-free regime.} $n^2p2^n\ll1$ (dashed line \(n^2 p_{\max} 2^n = 1\)), Costs equal noiseless baseline.
\item \textbf{Region II: Transitional regime.} $1\lesssim n^2p2^n \lesssim 10^2$, gradual increase with $n$.
\item \textbf{Region III: Noise-dominated regime.} $n^2p2^n\gg1$, rapid growth with clear separation by $p$.
\end{enumerate}
Log–log plots confirm polynomial growth in $n$ and divergence between noise levels only at sufficiently large $n$. This matches theoretical expectations.

\subsubsection{Pruning Mode Performance}
Figures~\ref{fig:QRAM_Benchmark_Total_Memory}(a,b) show that pruning preserves the same scaling patterns as full mode while reducing absolute costs. Ratios of noisy to noiseless memory, e.g. 2.01 ($p=10^{-5}$) and 11.93 ($p=10^{-4}$), agree with the predicted linear dependence,
\[
\frac{\text{Mem}_{\text{noisy}}}{\text{Mem}_{\text{noiseless}}} \approx 1 + p \cdot \mathrm{polylog}(N).
\]
Direct comparisons [Figs.~\ref{fig:QRAM_Benchmark_Total_Memory}(c,d)] show that pruning reduces runtime to $\sim$20\% and memory to $\sim$60\% of full-mode costs in the noise-dominated regime. These savings are largest exactly where resources are most demanding, demonstrating pruning’s effectiveness without altering the underlying physics. A detailed analysis of static vs. dynamic costs, small-$n$ behavior, and additional benchmark configurations is provided in Supplementary Information~\ref{Sec: Extended Analysis SI}.

Taken together, these benchmarks demonstrate that both address size $n$ and noise strength $\varepsilon$ contribute systematically to the overhead associated with noise handling. By confirming that our simulator reproduces the theoretically predicted scaling in both parameters, we establish its reliability across a wide operating range. This capability is crucial for exploring two distinct but equally relevant regimes: (i) large address sizes at low noise strengths, representing the asymptotic target of scalable QRAM research, and (ii) small address sizes at high noise strengths, reflecting the present reality of noisy devices with limited qubit counts. In both cases, our simulator provides a faithful tool for quantifying performance far beyond what existing state-vector approaches can access.

Building on this foundation, we next consider error filtration (EF). On the one hand, error filtration (EF) has been proven to be a practical, hardware-efficient method for extending QRAM capability without requiring full quantum error correction. On the other hand, EF is inherently a gate-based protocol, making it an ideal case study for demonstrating how our simulator integrates seamlessly with circuit-level quantum algorithms. We therefore now turn to EF as both a concrete application and a platform for probing the limits of QRAM error suppression at larger sizes and higher noise levels than previously accessible.
\begin{figure*}[ht]
    \centering
    \includegraphics[width=\textwidth]{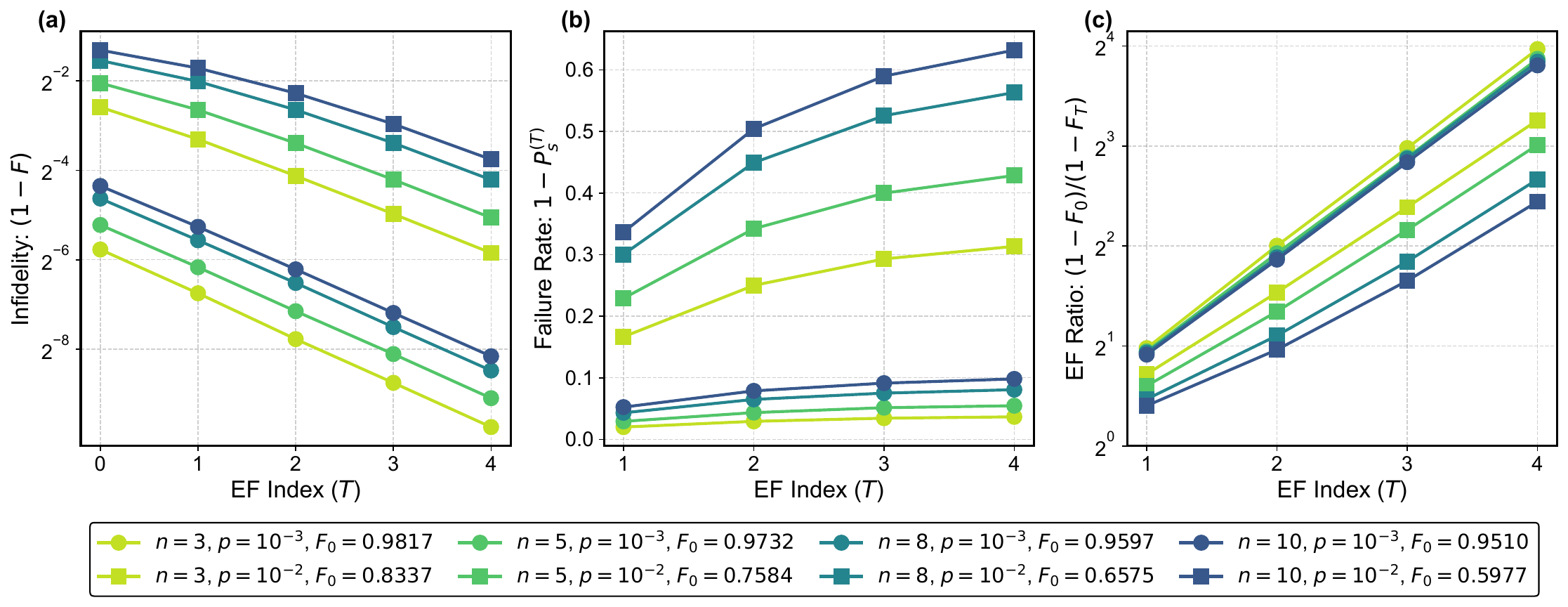}
    \caption{
    \textbf{Emergence of error filtration anomalies in BB QRAM simulations.}  \\
    \textbf{(a)} Log-scale infidelity \( 1 - F \) versus EF level \( T \), plotted for QRAM address sizes \( n = 3, 5, 8, 10 \) and depolarizing noise levels \( \varepsilon = 10^{-3}, 10^{-2} \). 
    Circle and square markers denote noise strength; colors denote QRAM size. 
    Ideal EF behavior predicts a slope of \(-1\), corresponding to infidelity halving with each repetition. 
    Deviations are visible at large \( n \) or stronger noise and the fidelity $F_0$ is included in the legend. \\
    \textbf{(b)} Post-selection failure probability \( 1 - P^{(T)}_S \), plotted across \( n \) and \( \varepsilon \), The failure probability for QRAM with high fidelity quickly plateau, showing consistent agreement with previous literature. And for the QRAM with moderate fidelity like 0.8 or less, the failure probability cannot be ignored as before.\\
    \textbf{(c)} Suppression ratio \( R_T = \frac{1 - F_0}{1 - F_T} \) versus \( T \), revealing the onset of EF anomalies. 
    While low-infidelity cases follow the ideal \( 2^T \) trend, stronger noise results in sub-exponential suppression and early saturation. 
    }
    \label{fig:Error_Filtration_Results}
\end{figure*}
\begin{figure*}[ht]
    \centering
    \includegraphics[width=\textwidth]{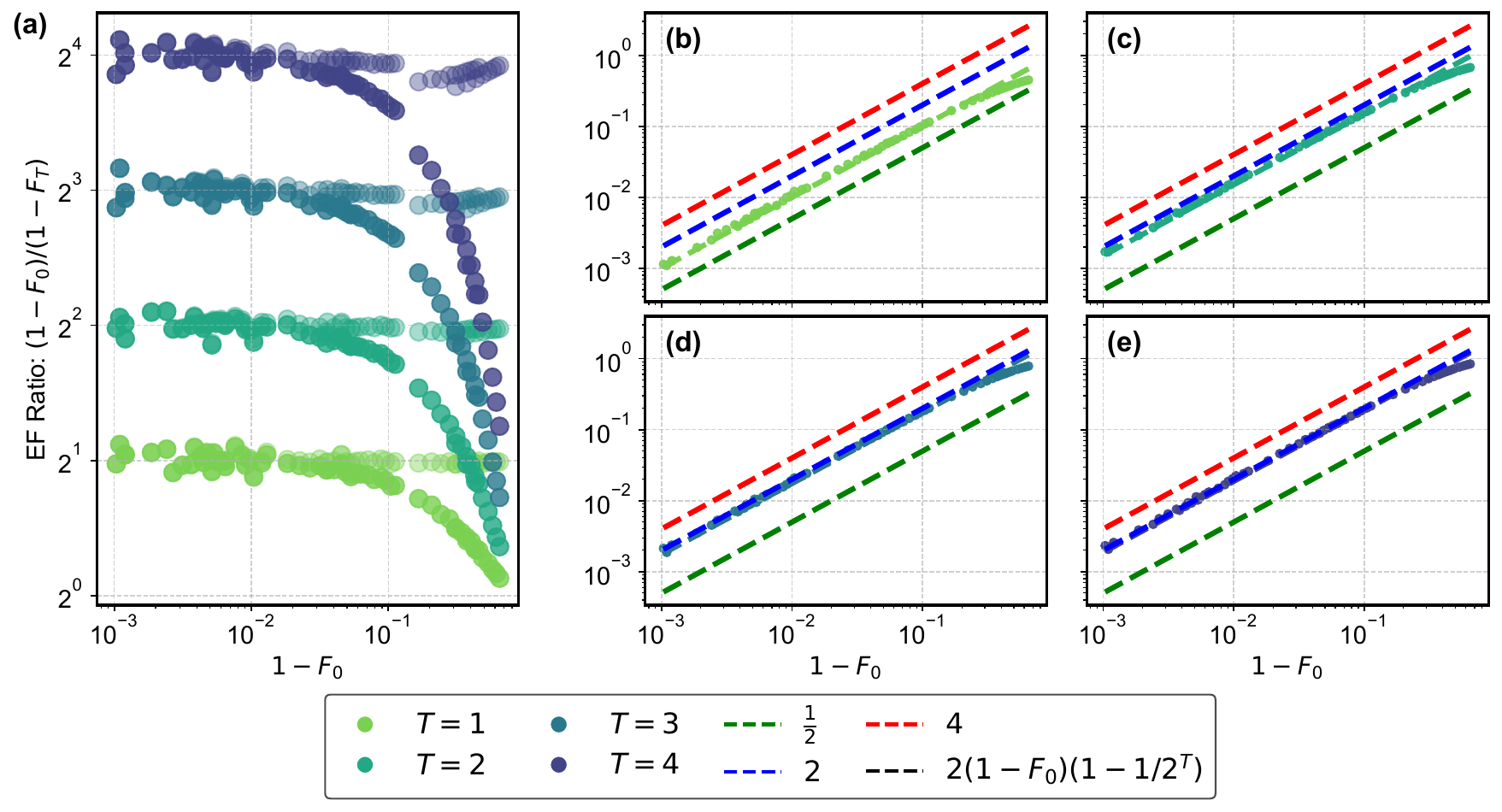}
    \caption{
    \textbf{EF suppression ratio and post-selection scaling in BB QRAM.}  \\
    \textbf{(a)} Numerical suppression ratio $\frac{1-F_0}{1-F_T}$ for $T=1,2,3,4$ over base infidelities $1-F_0 \in [10^{-5},10^{-2}]$ and address sizes $n=3$--$15$ with full-branch simulations.  
    Light markers show the same data divided by the post-selection probability $P_S^{(T)}$, demonstrating that deviations from the ideal $2^T$ scaling are largely explained by the neglected $P_S^{(T)}$ factor, particularly when $1-F_0 \gtrsim 0.1$. \\
    \textbf{(b--e)} Success probability $P_S^{(T)}$ versus base infidelity $1-F_0$, compared to the analytical bounds in Eq.~(\ref{eq:ps_bound_original}) and Eq.~(\ref{eq:ps_bound_refined}).  
    Red, blue, and green curves correspond to $4(1-F_0)$, $2(1-F_0)$, and $0.5(1-F_0)$ scaling trends, respectively.  
    The refined bound of Eq.~(\ref{eq:ps_bound_refined}) remains constant with $T$, confirming that exponential decay of $P_S^{(T)}$ does not occur in our BB QRAM simulations.
    }
    \label{fig:Ratio_Scaling_and_Bound}
\end{figure*}
\section{Error Suppression and Filtration\label{Sec: EF}}
Having described the EF protocol and its theoretical promise for suppressing noise in quantum circuits in Section~\ref{Sec: Preliminaries}, we now use our simulator to test EF performance in realistic QRAM systems. 
Our large-scale simulations, enabled by the scalability of the QRAM framework, allow us to go far beyond previously accessible regimes and directly probe EF behavior under increasing circuit depth and noise strength.
In addition, the general-purpose simulator that we integrate with the QRAM module has been independently validated under depolarizing noise and error filtration in Section~\ref{Sec: sparse simulator testing SI} of Supplementary Information, ensuring that EF performance is faithfully captured.
This capability reveals suppression anomalies that are invisible in purely asymptotic analyses.

\subsection{Emergence of Error Filtration Anomalies}

We quantify EF performance using the output infidelity \( 1 - F_T \) as a function of EF level \( T \). 
In the ideal regime, each EF level halves the infidelity, producing an exponential suppression ratio \( R_T \sim 2^T \) and a slope of \(-1\) on a logarithmic plot of infidelity vs.\ \( T \).  

To test this scaling, we conduct numerical experiments for address sizes \( n = 3, 5, 8, 10 \) under depolarizing noise with strengths \( \varepsilon = 10^{-3} \) and \( 10^{-2} \). 
For each setting, we generate 100 random input states and simulate 1000 shots per state. 
The results are shown in Fig.~\ref{fig:Error_Filtration_Results}.

As expected, systems with low noise and small \( n \) closely follow the ideal suppression trend. 
However, increasing noise strength or address size leads to noticeably weaker suppression, with slopes deviating from the \(-1\) benchmark and ratios saturating earlier than predicted.  
For instance, \( n=3 \) with \( \varepsilon=10^{-3} \) maintains \( 2^T \) scaling up to \( T=4 \), whereas with \( \varepsilon=10^{-2} \) the ratio saturates near \( 2^3 \) for \( T=4 \), despite fidelities still above 0.8.

These deviations are not explained by the standard small-\(\varepsilon\) EF formula, which predicts that non-ideal effects are second order in the base infidelity.  
Our results instead show that circuits with \( 1 - F_0 \gtrsim 0.1 \) rarely achieve the full \( 2^T \) suppression, and the discrepancy grows with \( T \).  
This points to a missing factor in the theory, motivating a refinement.

\subsection{Role of Post-Selection Probability in EF Performance}

EF post-selects the output state based on a set of ancilla qubits, yielding the normalized memory state
\begin{equation}
\rho^{(T)} = \frac{\tilde{\rho}^{(T)}}{P_S^{(T)}},
\end{equation}
where \( \tilde{\rho}^{(T)} \) is the unnormalized post-selected state and \( P_S^{(T)} \) is the probability of obtaining all ancillas in \(\ket{0}\).
Then, the EF of level $T$ infidelity is
\begin{equation}
(1-F)_T = 1 - \langle U\psi | \rho^{(T)} | U\psi \rangle.
\end{equation}

In our notation, the success probability \(P_S^{(T)}\) and unnormalized state \(\tilde{\rho}^{(T)}\) are given by  
\begin{align}
P^{(T)}_s 
&= \frac{1}{2^T}\nonumber\\
&\quad + \frac{1}{4^{T}} \sum_{i} \sum_{t=1}^{2^T} \sum_{q \ne t} 
\left( \langle \psi | K_{i_q}^\dagger K_{i_t} | \psi \rangle
\ \mathrm{Tr} \rho_\phi\, \overline{K_{i_q}^\dagger}\, \overline{K_{i_t}} \right),
\label{eq:ps_general}
\end{align}

\begin{align}
\tilde{\rho}^{(T)} 
&= \frac{1}{2^T} \, \mathcal{U}\big( |\psi\rangle \langle \psi| \big) \nonumber\\
&\quad + \frac{1}{4^T} \sum_{i} \sum_{t=1}^{2^T} \sum_{q \ne t} 
\left( K_{i_t} |\psi\rangle \langle \psi| K_{i_q}^\dagger 
\ \mathrm{Tr}\, \rho_\phi \, \overline{K_{i_q}^\dagger} \, \overline{K_{i_t}} \right),
\label{eq:rho_tilde_general}
\end{align}
where \(\overline{K}_{i_t}\) denotes the product of all Kraus operators except \(K_{i_t}\).

\paragraph*{Case \(T=1\).}
For clarity, we first consider \(T=1\). 
The success probability becomes
\begin{equation}
P^{(1)}_S = \frac{1}{2} \left[ 1 + \sum_{i,j} \mathrm{Tr}\!\left( K_i \rho_\psi K_j^\dagger \right) \, \mathrm{Tr}\!\left( K_j \rho_\phi K_i^\dagger \right) \right],
\end{equation}
and the unnormalized memory state is
\begin{equation}
\tilde{\rho}^{(1)} = \frac{1}{2} \left[ \mathcal{U}(\rho_\psi) + \sum_{i,j} K_i \rho_\psi K_j^\dagger \, \mathrm{Tr}\!\left( K_j \rho_\phi K_i^\dagger \right) \right].
\end{equation}
Up to \(\mathcal{O}(\varepsilon^2)\), these satisfy
\begin{equation}
\langle \Psi | \tilde{\rho}^{(1)} | \Psi \rangle 
= P_S^{(1)} - \frac{1}{2}(1 - F_0),
\end{equation}
where \(F_0\) is the base fidelity. Substituting into the definition of \((1-F)_1\) gives
\begin{equation}
(1-F)_1 = \frac{1-F_0}{2 P_S^{(1)}}.
\end{equation}
Thus, the suppression ratio becomes
\begin{equation}
\frac{1-F_0}{1-F_1} = 2 P_S^{(1)},
\end{equation}
revealing that the ideal suppression factor of \(2\) is reduced by the success probability.

\paragraph*{General \(T\).}
The above reasoning generalizes to
\begin{equation}
P_S^{(T)} - \langle \Psi | \tilde{\rho}^{(T)} | \Psi \rangle
= \frac{1}{2^T} (1 - F_0) + \mathcal{O}(\varepsilon^2),
\end{equation}
yielding
\begin{equation}
(1 - F)_{T}
= \frac{1-F_0}{2^T P_S^{(T)}} + \mathcal{O}(\varepsilon^2),
\end{equation}
so that the practical suppression ratio becomes
\begin{equation}
\frac{1-F_0}{1-F_T} = 2^T P_S^{(T)}.
\end{equation}
This matches our simulation results as shown in Fig.~\ref{fig:Ratio_Scaling_and_Bound}(a), where the ideal \(2^T\) scaling is recovered after dividing the measured ratio by \(P_S^{(T)}\). Complete analysis is in Section.~\ref{Sec: Derivation of EF SI} of the Supplementary Information.

\subsection{Refined Scaling of Post-Selection Probability}
In noise regimes with \( 1 - F_0 \gtrsim 0.1 \), the effectiveness of EF is determined jointly by the suppression factor \(2^T\) and the success probability \(P_S^{(T)}\).
A pessimistic, worst-case assumption—that any error \(K_{i>0}\) at any step causes rejection—yields the bound
\begin{equation}
P^{(T)}_{S} \ge 1 - 2^T\varepsilon + \mathcal{O}(\varepsilon^2),
\label{eq:ps_worst_case}
\end{equation}
which predicts exponential decay in \(P_S^{(T)}\) with \(T\) and would make high-\(T\) EF impractical.

The original EF analysis~\cite{Gideon_EFiltration_2023} established the rigorous lower bound
\begin{equation}
P_S^{(T)} \ge 1 - 4\varepsilon + \frac{\varepsilon}{2^T},
\label{eq:ps_bound_original}
\end{equation}
valid under certain favorable conditions.  
By further assuming identical inputs for the memory and ancilla registers, we derive the improved bound
\begin{equation}
P_S^{(T)} \ge 1 - 2\varepsilon,
\label{eq:ps_bound_refined}
\end{equation}
which is \emph{independent} of \(T\).  
This removes the exponential penalty from Eq.~\eqref{eq:ps_worst_case}, ensuring that EF retains its exponential noise-suppression advantage at large $T$ and numerical evidence is all presented in Fig.~\ref{fig:Ratio_Scaling_and_Bound}(b-e) and further analysis is in Section.~\ref{Subsec: General T Analysis} (c) of the Supplementary Information.

The refined lower bound in Eq.~(\ref{eq:ps_bound_refined}) has direct implications for the range of QRAM sizes where EF remains beneficial.  
If we define a \emph{progressive EF condition}—that EF at level $T$ must outperform all lower levels, specifically satisfying
\[
2^T P_S^{(T)} \ge 2^{T-1},
\]
then $P_S^{(T)} \ge 0.5$.  
From Eq.~(\ref{eq:ps_bound_refined}), this condition requires the base infidelity $\varepsilon$ to obey
\[
\varepsilon \le \frac{1}{2C_{\mathrm{bound}}},
\]
where $C_{\mathrm{bound}}$ is the constant in the $T$-independent lower bound.  
Our refinement halves $C_{\mathrm{bound}}$ from $4$ to $2$, thereby doubling the tolerable $\varepsilon$ (e.g., from $0.125$ to $0.25$).  
\begin{figure}[ht]
    \centering
    \includegraphics[width=\columnwidth]{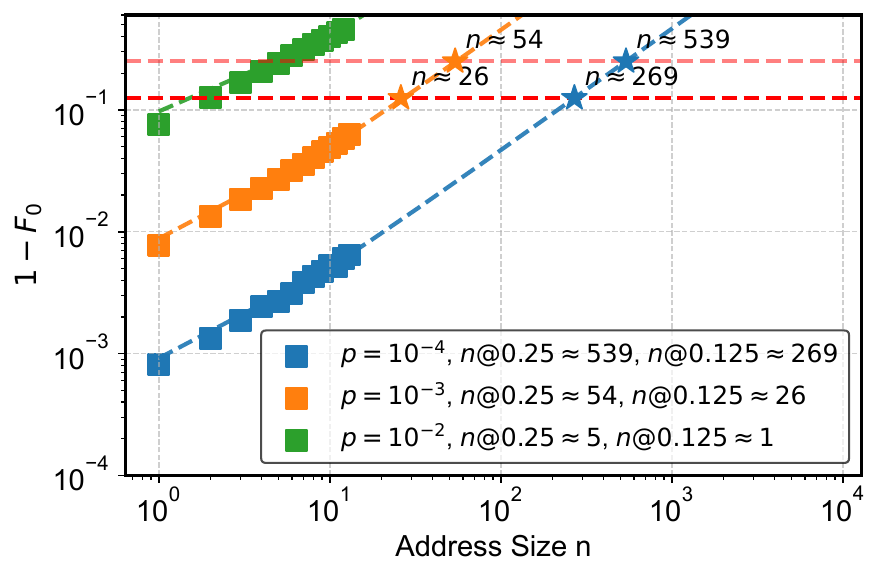}
    \caption{
    \textbf{Estimated maximum feasible BB QRAM size under progressive EF.}  
    Simulated base infidelity $(1-F_0)$ for noiseless BB QRAM as a function of address size $n$ (dots) with a power-law fit $n^{1.9}$ (solid line).  
    Horizontal dashed lines correspond to the maximum tolerable base infidelity $\varepsilon_{\mathrm{max}}$ required by the \emph{progressive EF condition} $2^T P_S^{(T)} \ge 2^{T-1}$, where $P_S^{(T)}$ is bounded below by Eq.~(\ref{eq:ps_bound_original}) (red) and the refined bound Eq.~(\ref{eq:ps_bound_refined}) (blue).  
    Intersection points (vertical dashed lines) indicate the largest QRAM sizes where EF can still guarantee progressive improvement.  
    The refined bound doubles $\varepsilon_{\mathrm{max}}$ (e.g., from $0.125$ to $0.25$), extending the feasible range of $n$ by more than a factor of two.
    }
     \label{fig:Rough_Estimate_on_Possible_Size}
\end{figure}
To quantify the impact on achievable QRAM sizes, we fit the simulated noiseless infidelity growth for $n=1$ to $15$ with a power law $n^{1.9}$, as shown in Fig.~\ref{fig:Rough_Estimate_on_Possible_Size}.  
The intersection of this scaling curve with the horizontal threshold lines set by the base-infidelity tolerance determines the maximum feasible QRAM size.  
Under the refined bound, the allowed range of $n$ extends by more than a factor of two compared to the original bound, significantly expanding the experimentally relevant regime for EF-enhanced BB QRAM.

Our findings illustrate how the performance of EF in BB QRAM depends not only on the expected exponential suppression factor but also on the success probability of post-selection, leading us to refine theoretical bounds and extend the parameter regimes where EF is practically beneficial. More broadly, these results showcase the value of our simulator as an end-to-end analysis tool. It validates known asymptotic scaling, uncovers hidden anomalies, quantifies trade-offs between size and noise strength, and demonstrates seamless integration with gate-level protocols such as EF. In doing so, it establishes a foundation for future studies of QRAM fault tolerance and for assessing the true algorithmic value of QRAM as a quantum resource.

\section{Conclusions and Outlook}
\label{sec: Conclusions}

We have presented a scalable classical simulation framework for bucket-brigade QRAM that combines sparse state encoding with a noise-aware pruning algorithm.  
This approach reduces simulation cost from exponential in the number of memory cells to polynomial in practice, even in the presence of noise, and enables seamless integration of QRAM modules into larger gate-based quantum algorithms.

Using this tool, we performed the first large-scale, state-level study of the error filtration (EF) protocol in realistic noisy QRAMs.  
Our simulations extend far beyond the size and noise regimes accessible to prior work, uncovering suppression anomalies that are invisible in purely asymptotic analyses.  
By explicitly incorporating the post-selection probability into EF theory, we refined the analytical suppression model, obtaining a near-deterministic relation between base infidelity and achievable suppression ratio.  
This refinement allows us to delineate both the practical range of EF operation and the maximum QRAM sizes for which EF yields progressive improvement—thereby providing concrete, quantitative guidance for near-term QRAM deployment.

More broadly, our simulator acts as an efficient QRAM research engine. It validates theoretical predictions where first-order approximations hold, exposes deviations that emerge at larger scales, and supplies the “fine print” necessary to judge QRAM as a realistic computational resource. Its ability to provide full quantum state access, coupled with scalable performance under realistic fault conditions, makes it a powerful tool for bridging the gap between abstract QRAM models and experimental realization. Looking ahead, extending these methods to richer noise models and broader classes of algorithms will enable comprehensive end-to-end analyses of QRAM performance, clarifying both its limitations and its potential as a cornerstone of quantum computing.

\section{Acknowledgement\label{sec: Acknowledgement}}
This work has been supported by the National Key Research and Development Program of China (Grant Nos. 2023YFB4502500 and 2024YFB4504100), the National Natural Science Foundation of China (Grant No. 12404564), and the Anhui Province Science and Technology Innovation (Grant Nos. 202423s06050001 and 202423r06050002).

\bibliography{ref.bib}

\clearpage
\setcounter{table}{0}
\renewcommand{\thetable}{S\arabic{table}}%
\setcounter{figure}{0}
\renewcommand{\thefigure}{S\arabic{figure}}%
\setcounter{section}{0}
\setcounter{equation}{0}
\renewcommand{\theequation}{S\arabic{equation}}%

\onecolumngrid

\begin{center}
    {\large \bf Supplementary Information: \\ Refined Criteria for QRAM Error Suppression via Efficient Large-Scale QRAM Simulator}
    \vspace{0.3cm}
\end{center}

\section{Introduction to Bucket-Brigade QRAM\label{sec: Intro SI}}
\begin{figure}[ht]
    \centering
    \includegraphics[width=0.4\columnwidth]{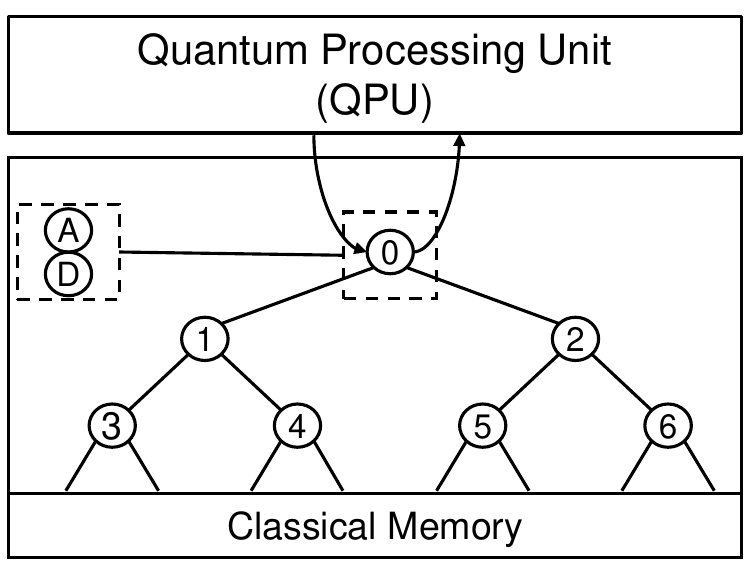}
    \caption{\textbf{Schematic of the bucket-brigade QRAM.}
    The QRAM serves as the interface between a quantum processing unit (QPU) and classical memory, enabling coherent access to $N=2^n$ memory cells. Address and data qubits are routed through a binary tree of ancillary nodes, indexed from the root (0) down to depth 3 in this example. Nodes are labeled in top-down, left-to-right order, beginning with the root at index 0. Layer $i$ contains $2^i$ nodes, with indices starting from $2^i - 1$. 
    }
    \label{fig: QPU-QRAM Interface}
\end{figure}

This section provides a comprehensive overview of the bucket-brigade quantum random access memory (BB-QRAM), a structured framework for coherent quantum memory access. In classical RAM systems, each memory location is indexed by a unique binary address, and a query simply involves providing the address as input and reading out the corresponding value. In contrast, QRAM enables access in quantum superposition, allowing simultaneous queries across multiple addresses—a capability that is central to many quantum algorithms. Practical implementations of QRAM, such as those proposed in Refs.\cite{QRAM_Parallel_Chen_2023, QRAM_Archi_Seth_2008}, typically consist of two main components: a data bus and an ancillary binary tree, as illustrated in Fig.\ref{fig: QPU-QRAM Interface}. 
The data bus serves as the communication interface between the QRAM module and other components of a quantum computing architecture, including quantum processing units (QPUs)~\cite{liu2023quantummemorymissingpiece, YongShan_QRAMSystem_2023}. Both address qubits and data qubits are routed into the binary tree via this bus, and at the terminal layer, the tree connects to $N$ classical memory cells.

In this work, we adopt the query protocol introduced in Ref.~\cite{QRAM_Parallel_Chen_2023}, which decomposes a QRAM operation into three conceptual stages. First, during address setting, the address information injected through the bus establishes routing paths from the root of the binary tree to the appropriate leaf nodes. These paths determine the traversal of data qubits in subsequent steps. Second, in the data fetching stage, the data qubits are propagated along the established paths to the corresponding memory cells, where they interact with the stored classical values. Finally, the uncomputing stage routes the data qubits back to the data bus by reversing all operations performed during address setting and data fetching. This disentangles the ancillary tree from the bus and ensures that no residual entanglement remains in the system.

In this architecture, the depth of a QRAM query scales as $\mathcal{O}(\log N)$, an attractive feature for quantum algorithms seeking exponential speedups over classical methods. However, this advantage comes with a substantial hardware cost: each query requires $\mathcal{O}(N)$ ancillary qubits and operations. The exponential demand for qubits and gates also makes classical simulation of QRAMs extremely challenging—one of the key difficulties addressed in this work.

\section{Quantum Circuit Simulators \label{Sec: Sparse SI}}
Current quantum circuit simulators are generally based on one of two paradigms: Schr\"{o}dinger-based (statevector) simulation or Feynman-based path summation. 
Each approach offers distinct trade-offs in terms of computational complexity, memory requirements, and scalability.

In Schr\"{o}dinger-based (Statevector) approach, the quantum state of an $n$-qubit system is explicitly represented as a statevector with $2^n$ complex amplitudes. 
The computational complexity for a circuit with $m$ gates is $O(m 2^n)$, since each gate operation involves updating $O(2^n)$ amplitudes. It is conceptually straightforward and provides exact results, but the memory cost grows exponentially with $n$. 
For example, simulating a 50-qubit system requires more than one petabyte of memory~\cite{qubits_45}. 
To mitigate this, sparse representations have been developed.
Many quantum states remain sparse during their evolution, especially in structured algorithms, allowing simulators to store and manipulate only the nonzero amplitudes. Leveraging this property, sparse frameworks have demonstrated simulations of systems with over 100 qubits using modest classical resources~\cite{Samuel_SparseSimulator_2022}.

The Feynman approach, by contrast, computes the evolution of a circuit by summing over paths rather than storing a full statevector. A Feynman path is defined as a sequence of basis-state transitions across $m$ layers of gates, and the output amplitude is obtained by summing contributions from all valid paths. In the worst case, the number of paths scales as $2^{n(m-1)}$, leading to exponential complexity. However, this method has notable advantages: each path requires only $O(m)$ memory, evaluations are naturally parallelizable, and paths with zero amplitude can be pruned. The efficiency of this approach depends critically on the gate set, since certain gates either preserve or expand the path structure.

This observation motivates a useful classification of gates into non-branching and branching types, based on their action on computational basis states. This distinction plays a central role in the performance of Feynman-based simulation techniques.

\paragraph{Non-branching gates:}
A non-branching gate is one whose matrix representation has at most one nonzero entry per row, meaning it does not generate superpositions but only permutes or phases basis states.
Common examples include the identity gate ($\GateI$), Pauli gates ($\GateX$, $\GateY$, $\GateZ$), controlled-Z gate ($\GateCZ$), as well as multi-qubit gates like the Toffoli and Fredkin gates. 
For instance, the $\GateCZ$ gate simply adds a phase to $|11\rangle$ while leaving the computational basis structure intact.

\paragraph{Branching gates:}
In contrast, branching gates map a single basis state to multiple outcomes, thereby generating superpositions and new paths.
A typical example is the Hadamard gate ($\GateH$) , which transforms:
\begin{equation}
    |0\rangle \to \frac{1}{\sqrt{2}}(|0\rangle + |1\rangle), \quad |1\rangle \to \frac{1}{\sqrt{2}}(|0\rangle - |1\rangle).
\end{equation}
Layers of branching gates can therefore cause exponential growth in the number of active paths, while layers of purely non-branching gates leave the path count unchanged.

This classification plays a central role in optimizing Feynman-style simulation. Circuits dominated by non-branching gates avoid exponential path growth, allowing both memory and runtime costs to remain manageable. The bucket-brigade QRAM architecture is particularly favorable in this respect, as its query protocol is built almost entirely from non-branching operations such as Fredkin and Pauli gates. Consequently, the number of computational paths remains constant throughout a query, even when realistic noise models such as Pauli channels are included. To capture this property more formally, we introduce the notion of a branch, defined as the number of nonzero-amplitude basis states in the address register at the start of the QRAM query. Since the complexity of Feynman-style simulation scales directly with the number of branches rather than the full memory size, the overall cost is determined by the structure of the input superposition, not the exponential Hilbert-space dimension. This observation underpins the scalability of our framework and explains why BB-QRAM circuits remain tractable to simulate even in noisy, large-scale instances.

\section{Noise channels and simulation\label{Sec: Noise Channels SI}}
We apply our simulation algorithm to compute query infidelity in QRAM circuits under qutrit error channels, following the definitions established in previous works~\cite{ramzan2011effectflippingnoiseentanglement}.

The qutrit operators are defined as:
$$A_1=\begin{pmatrix}
0  & 1 & 0\\
1  &  0&0 \\
 0 &0  & 1
\end{pmatrix},
A_2=\begin{pmatrix}
1 & 0 & 0 \\
0 & \omega & 0 \\
0 & 0 & \omega^{2}
\end{pmatrix}.
$$
where the basis is $\{|W\rangle,|0\rangle,|1\rangle\}$, and $\omega=e^{i2\pi/3}$.

The qutrit error channels are defined using their Kraus decomposition $\{K_0, K_1, \cdots, K_{i}\}$  as follows:
$$
\begin{aligned}
\text { Depolarizing } &=\left\{\sqrt{1-\varepsilon} I, \sqrt{\frac{\varepsilon}{8}} A_{1}, \sqrt{\frac{\varepsilon}{8}} A_{2}, \sqrt{\frac{\varepsilon}{8}} A_{1}^{2}, \sqrt{\frac{\varepsilon}{8}} A_{2}^{2}, \sqrt{\frac{\varepsilon}{8}} A_{1} A_{2}, \sqrt{\frac{\varepsilon}{8}} A_{1}^{2} A_{2}, \sqrt{\frac{\varepsilon}{8}} A_{1} A_{2}^{2}, \sqrt{\frac{\varepsilon}{8}} A_{1}^{2} A_{2}^{2}\right\} \\
\text { Damping } &=\left\{|W\rangle\langle W|+\sqrt{1-\varepsilon}(|0\rangle\langle 0|+| 1\rangle\langle 1|), \sqrt{\varepsilon}| W\rangle\langle 0|, \sqrt{\varepsilon}| W\rangle\langle 1|\right\} \\
\text { Heating } &=\left\{|0\rangle\langle 0|+| 1\rangle\langle 1|+\sqrt{1-\varepsilon}| W\rangle\langle W|, \sqrt{\frac{\varepsilon}{2}}| 0\rangle\langle W|, \sqrt{\frac{\varepsilon}{2}}| 1\rangle\langle W|\right\} .
\end{aligned}
$$
These error models allow us to characterize the effects of depolarization, damping, and heating on QRAM performance, providing insight into the robustness of QRAM queries under realistic noise conditions.

Among the different types of channels, the mixed-unitary channel is of particular interest. In this case, each Kraus operator is proportional to a unitary, i.e., $K_i=\sqrt{p_i} U_i$ for all $i$, as in bit-flip errors or the general depolarizing channel~\cite{Takahashi_ClassicalDepol_2021}.
Simulation of mixed-unitary noise is relatively straightforward, since the output probability is independent of the input state. The simulator randomly selects the operator $U_i$ according to its probability $p_i$, and otherwise proceeds exactly as in a noiseless circuit. Before each shot, noise locations are sampled from the circuit according to the specified error rate.

In contrast, non-unitary channels behave differently. In a mixed-unitary channel the system often remains unchanged, whereas in a non-unitary channel the state is always biased away from its original form. Our simulation algorithm therefore designates the high-probability Kraus operator as the “correct” biased operator, applied by default, while the remaining probability mass is treated as noise spots that require explicit handling. If a noise spot is sampled, the simulator performs a quasi-measurement to resolve the effect.
To make this concrete, we consider the qubit amplitude damping channel. One of its noise operators is $E_1=\sqrt{\gamma}|0\rangle\langle 1|$, and the probability of its occurrence for an input state $|\psi\rangle$ is
$$
    p_1 = \langle \psi| E_1^\dagger E_1 |\psi \rangle
        = \gamma \langle\psi |1\rangle\langle 1|\psi\rangle
        = \gamma p_{|1\rangle},
$$
is the probability of finding the system in $|1\rangle$. The corresponding output state is
$$
\rho = p_0|\tilde{\psi}_0\rangle\langle\tilde{\psi}_0| + p_1|\tilde{\psi}_1\rangle\langle\tilde{\psi}_1|,
$$
where $|\tilde{\psi}_0\rangle = \tfrac{E_0}{\sqrt{p_0}}|\psi\rangle$ describes the normalized state when no error occurs, and $|\tilde{\psi}_1\rangle = \tfrac{1}{\sqrt{p_1}}|0\rangle\langle 1|\psi\rangle$ is the normalized state after an amplitude damping event.

Since $p_0=(1-\gamma)+\gamma p_{|0\rangle}$, we can further decompose the output state as:
$$
\rho = (1-\gamma)|\tilde{\psi}_0\rangle\langle\tilde{\psi}_0| + \gamma\left(
p_{|0\rangle}|\tilde{\psi}_0\rangle\langle\tilde{\psi}_0|+p_{|1\rangle}|\tilde{\psi}_1\rangle\langle\tilde{\psi}_1|\right).
$$
This formulation makes amplitude damping simulatable in essentially the same manner as unitary noise. For each simulation shot, noise spots are sampled with probability $\gamma$, and whenever a spot is encountered, the simulator performs a quasi-measurement to determine whether the qubit is in $|0\rangle$ or $|1\rangle$, applying $E_0$ or $E_1$ accordingly. In the absence of noise, $E_0$ is applied by default.

\section{Extended Benchmark Analysis \label{Sec: Extended Analysis SI}}
Here we provide additional details on the benchmark methodology used in Section~\ref{Sec: Method} of the main text.

All simulations were implemented in C++ following the algorithms of Section~\ref{Sec: Method} and executed on a 128-core server equipped with two 3.50~GHz Intel Xeon Platinum~8369B CPUs (64~cores each) and 512~GB of available memory.

To isolate the impact of noise, we separate total resource usage into two components:
\begin{itemize}
    \item \textbf{Static cost:} Baseline time and memory required to initialize the simulator and build its data structures, measured in the noiseless case with different branch sizes.
    \item \textbf{Dynamic cost:} Additional time and memory consumed during the simulation itself, relative to the static baseline. This includes the cost of processing multiple branches and, under noise, applying noise operators to unreliable branches.
\end{itemize}
The \emph{dynamic simulation cost} is therefore defined as
\[
\Delta\text{Cost} = \text{Cost}_{\text{noisy}} - \text{Cost}_{\text{noiseless}},
\]
measured separately for time and memory.  
From our theoretical analysis, the noisy dynamic cost should be dominated by the number of noisy branches, scaling as \(\mathcal{O}(n^2 p \, 2^n)\), where \(n\) is the address size and \(p\) the noise strength.  
Thus, the scaling of \(\Delta\text{Cost}\) with \((n,p)\) should mirror that of the infidelity.

\subsection{Static Cost Baseline}

\paragraph{Static Time.}  
In the noiseless case, the static runtime---measured with noise disabled and fixed branch size---is largely independent of the address size \(n\).  
Fig.~\ref{fig:QRAM_FullMode_Benchmark}(a) shows the measured baseline time for branch sizes \(2^0\), \(2^5\), \(2^{10}\), and \(2^{15}\).  
For small branch sizes (\(2^0\), \(2^5\), \(2^{10}\)), the runtime is on the order of \(10^{-1}\)–\(10^0\) ms and remains essentially flat for \(n \gtrsim 10\).  
At very small \(n\), runtimes approach the measurement resolution, producing visible scatter.  
For the largest branch size tested (\(2^{15}\)), the runtime is proportionally larger, but still exhibits no appreciable growth with \(n\).

These results confirm that the noiseless simulation cost depends almost entirely on branch size---i.e., the number of nonzero amplitudes in the sparse encoding---and is insensitive to the total number of address qubits.

\paragraph{Static Memory.}  
In the noiseless case, the static memory cost has two distinct contributions:  
(i) memory for storing branch states in the sparse encoding, and  
(ii) classical memory for holding the QRAM’s target data values for all \(2^n\) addresses.  
Fig.~\ref{fig:QRAM_FullMode_Benchmark}(b) shows the measured baseline memory usage for branch sizes \(2^0\), \(2^5\), \(2^{10}\), and \(2^{15}\).  
For small \(n\), branch storage dominates, giving a flat profile across different address sizes.  
As \(n\) increases, the classical data term scales as \(\mathcal{O}(2^n)\) and eventually overtakes branch storage, causing curves for different branch sizes to converge and exhibit the expected exponential growth.

While the classical data term is exponential in \(n\), this is unavoidable: faithfully emulating a QRAM query requires explicit instantiation of the classical memory cells retrieved by the quantum bus.  
Crucially, for fixed branch size, the cost remains flat over a wide range of \(n\) before the classical term dominates, demonstrating that our sparse encoding cleanly separates branch-related memory from classical data cost.

\subsection{Noise-Induced Overhead}

\paragraph{Dynamic Cost.}  
Figures~\ref{fig:QRAM_FullMode_Benchmark}(c,d) and Figures~\ref{fig:QRAM_Benchmark_Total_Memory}(a,b) show the dynamic simulation cost---the additional runtime and memory overhead relative to the noiseless baseline---for multiple noise configurations.  
Across all cases, the growth with \(n\) and \(\varepsilon\) follows our theoretical prediction: given a noise rate \(\varepsilon\), the expected number of faulty nodes scales as \(N \varepsilon\), where \(N = \mathcal{O}(2^n)\) is the total number of nodes in the BB QRAM binary tree.  
Because a single fault affects only the branches routed through its subtree, the total number of affected branches scales as
\[
\mathcal{O}\!\left(N \varepsilon \cdot \mathrm{polylog}(N)\right),
\]
there enabling the infidelity which is the number of faulty branches over total branches to $\mathcal{O}\!\left(\varepsilon \cdot \mathrm{polylog}(N)\right)$.
This restricted fault-propagation pattern is a direct consequence of the binary tree structure, and underpins the error resilience of BB QRAM compared to fully connected quantum memories.

For interpretation, we divide the \((n,p)\) space into three regimes:
\begin{enumerate}
    \item \textbf{Region I: Noise-free regime.}  
    \(n^2 p \, 2^n \ll 1\) (dashed line \(n^2 p_{\max} 2^n = 1\)).  
    The expected number of faulty branches is \(<1\), so noisy simulations match noiseless costs. Runtime per branch and total memory remain flat.
    \item \textbf{Region II: Transitional regime.}  
    \(1 \lesssim n^2 p \, 2^n \lesssim 256\).  
    Faulty branches grow from \(\mathcal{O}(1)\) to \(\mathcal{O}(10^2)\), producing gradual increases in per-branch load time and total memory.  
    Higher \(p\) values (e.g., \(10^{-4}\)) exhibit visible slope changes in log--log plots, though differences between noise levels remain modest.
    \item \textbf{Region III: Noise-dominated regime.}  
    \(n^2 p \, 2^n \gg 1\).  
    Most branches become unreliable, and cost grows significantly with both \(n\) and \(p\).  
    Runtime curves separate clearly by noise strength, with approximately polynomial dependence on \(n\) at fixed \(p\).
\end{enumerate}

In both figures, dynamic time and memory scale as straight lines in log--log plots, indicating polynomial growth in \(n\).  
For small \(n\), curves for different noise configurations nearly coincide, consistent with the noise-free regime where \(n^2 p 2^n \ll 1\).  
Divergence between noise configurations appears for \(n \gtrsim 14\), marking the onset of the transitional regime where the cumulative impact of \(n^2 p 2^n\) becomes non-negligible.  
For larger \(n\), the separation between curves becomes pronounced, matching the noise-dominated regime in which the number of noisy branches grows rapidly with \(n\) and \(p\).  
These trends confirm that the simulator accurately reproduces the theoretically expected scaling of noise-induced resource overhead.

\paragraph{Pruning Mode Performance.} 
In Figures~\ref{fig:QRAM_Benchmark_Total_Memory}(a,b), these three regions are explicitly indicated by shaded backgrounds: blue for Region~I (noise-negligible), yellow for Region~II (transitional), and red for Region~III (noise-dominated).  
The preservation of the same scaling patterns seen in Figures~\ref{fig:QRAM_FullMode_Benchmark}(c,d), with pruning enabled, confirms that the algorithmic speedups in pruning mode do not alter the underlying physical scaling of noise-induced resource overhead. The scaling behavior mirrors both of that in full mode and pruning mode: costs are flat in Region~I, grow gradually in Region~II, and rise sharply in Region~III with clear separation by noise level.  
Moreover, pruning substantially reduces both runtime and memory across all regimes,

Fig.~\ref{fig:QRAM_Benchmark_Total_Memory}(c,d) compare pruning-mode costs directly to full-mode costs by plotting the ratio of runtime and memory usage, respectively.  
In the noise-dominated regime at large \(n\), pruning reduces runtime to as low as \(\sim 20\%\) of the full-mode cost [Fig.~\ref{fig:QRAM_Benchmark_Total_Memory}(c)], and memory usage to \(\sim 60\%\) [Fig.~\ref{fig:QRAM_Benchmark_Total_Memory}(d)].  
These savings are particularly significant given that they occur in the most resource-intensive region of the parameter space, where the number of branches is large and noise effects are strongest.  
The order-of-magnitude reduction factors over the tested noise configurations demonstrate that pruning provides consistent performance gains without altering the underlying scaling behavior established in the full-mode benchmarks.

\section{Sparse Simulator under Depolarizing Noise with Error Filtration \label{Sec: sparse simulator testing SI}}

This project relies primarily on two custom simulators rather than external packages, making validation against analytically tractable models essential. To this end, we benchmarked the sparse simulator by incorporating depolarizing noise and applying error filtration (EF), testing these components both separately and in combination.

We begin with a baseline test of depolarizing noise alone, using the state fidelity as a figure of merit. For simplicity, we consider the identity operation, a special case of a unitary transformation, combined with depolarizing noise. The fidelity between the ideal and noisy outputs is defined as
\begin{equation}
    \label{eqn: state fidelity}
    \Tr(U \ket{\psi} \bra{\psi} U^{\dagger} \Tilde{U} (\ket{\psi} \bra{\psi})),
\end{equation}
where $U$ represents the ideal operation and $\widetilde{U}(\cdot)$ denotes the noisy channel. For the $N_r$-qubit depolarizing channel, the Kraus form is
\begin{equation}
    \label{eqn: depolarizing noise channel}
    \sum^{4^{N_{r}}-1}_{i=0} P_i \rho P_i^{\dagger} = (1-p) \rho + \frac{p}{4^{N_{r}}-1}\sum^{4^{N_{r}}-1}_{i=1} P_i \rho P_i^{\dagger},
\end{equation}
with single-qubit error probability $p$. The theoretical fidelity in this case is $(1 - 2p/3)^{N{r}}$.

\begin{figure}[ht]
    \centering
    \includegraphics[width=0.6\columnwidth]{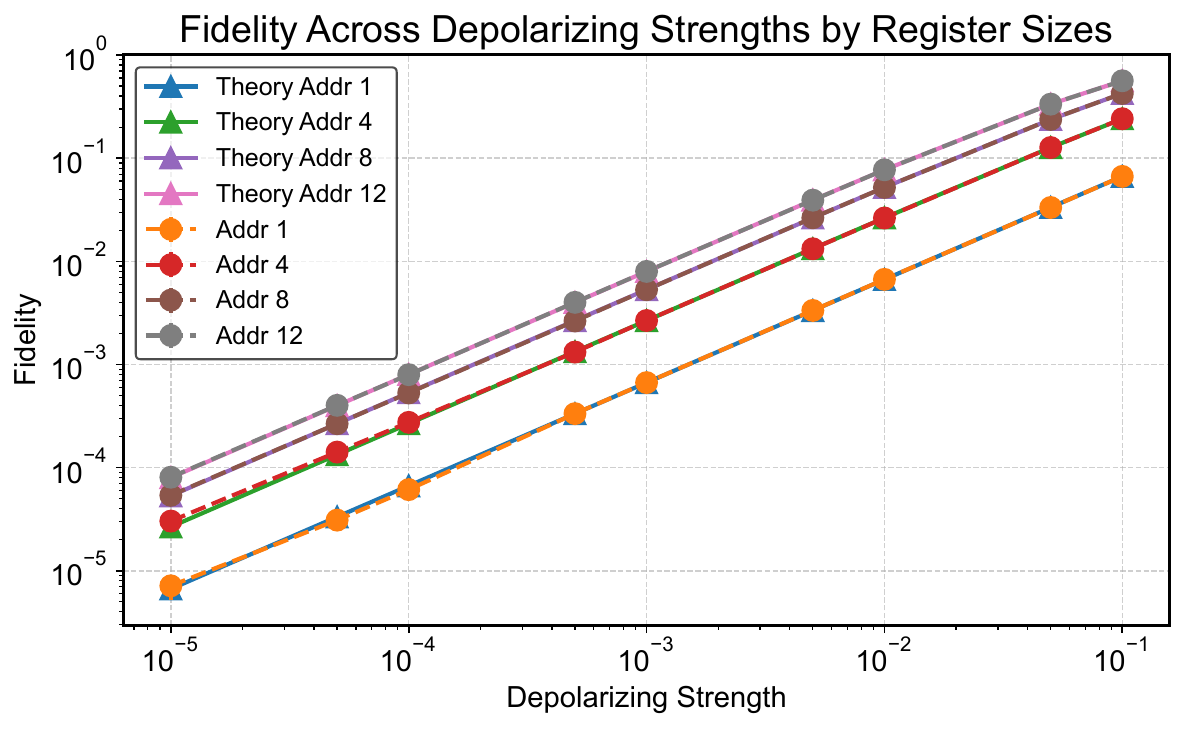}
    \caption{Identity channel with depolarizing noise. Simulation results closely follow the theoretical fidelity decay $(1 - 2p/3)^{N_r}$.}
    \label{fig: Fidelity across register size vs theoretical values}
\end{figure}

Simulations on registers of different sizes ($N_r = 1, 4, 8, 12$) reproduced the expected exponential fidelity decay, as shown in Fig.~\ref{fig: Fidelity across register size vs theoretical values}. The close agreement confirms the correct implementation of depolarizing noise in the sparse simulator.

\begin{figure}[ht]
    \centering
    \includegraphics[width=\columnwidth]{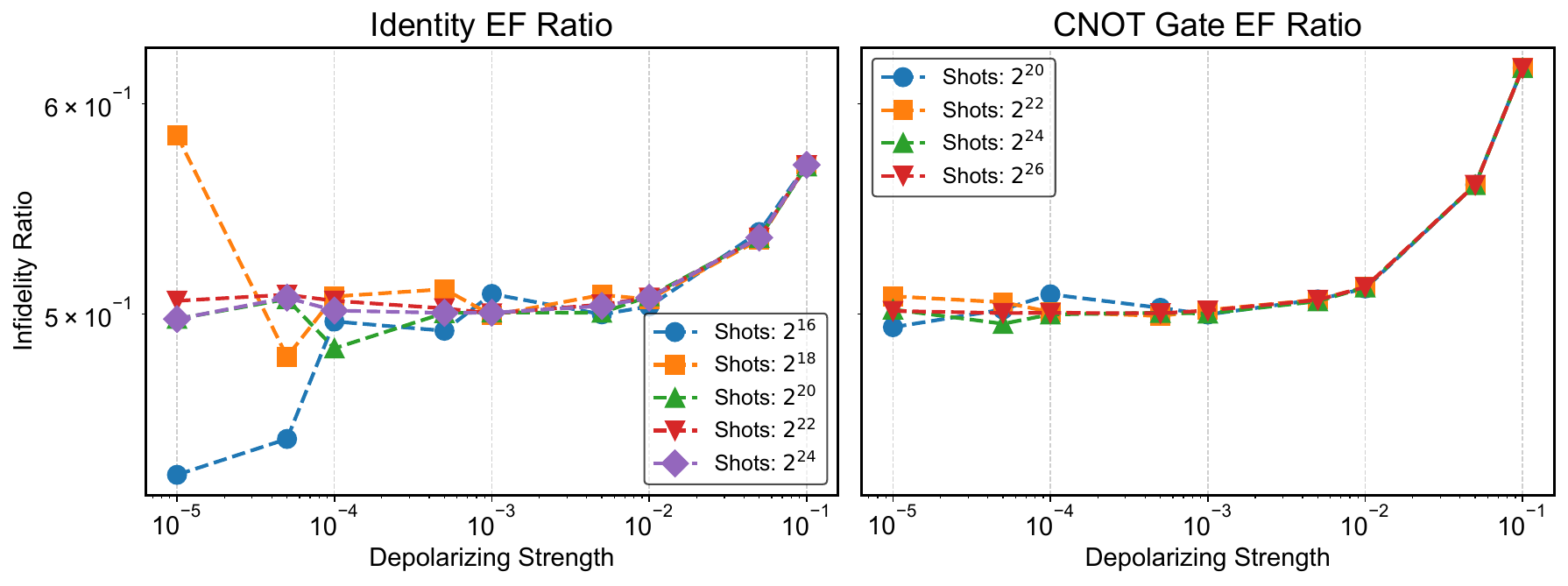}
    \caption{Error filtration applied to identity and CNOT operations under depolarizing noise. In the weak-noise limit, the infidelity ratio approaches $1/2$, as expected.}
    \label{fig:Different channels with depolarizing noise with error filtration}
\end{figure}

Next, we validated the integration of EF with depolarizing noise for both the identity and CNOT operations. As illustrated in Fig.~\ref{fig:Different channels with depolarizing noise with error filtration}, the ratio of post-filtration to pre-filtration infidelity converges to $1/2$ in the weak-noise limit, consistent with theory. At larger noise strengths, the ratio deviates, reproducing the same anomaly described in the main text. These results demonstrate that EF is faithfully implemented in the sparse simulator.

\begin{figure}[ht]
\centering
\includegraphics[width=0.6\columnwidth]{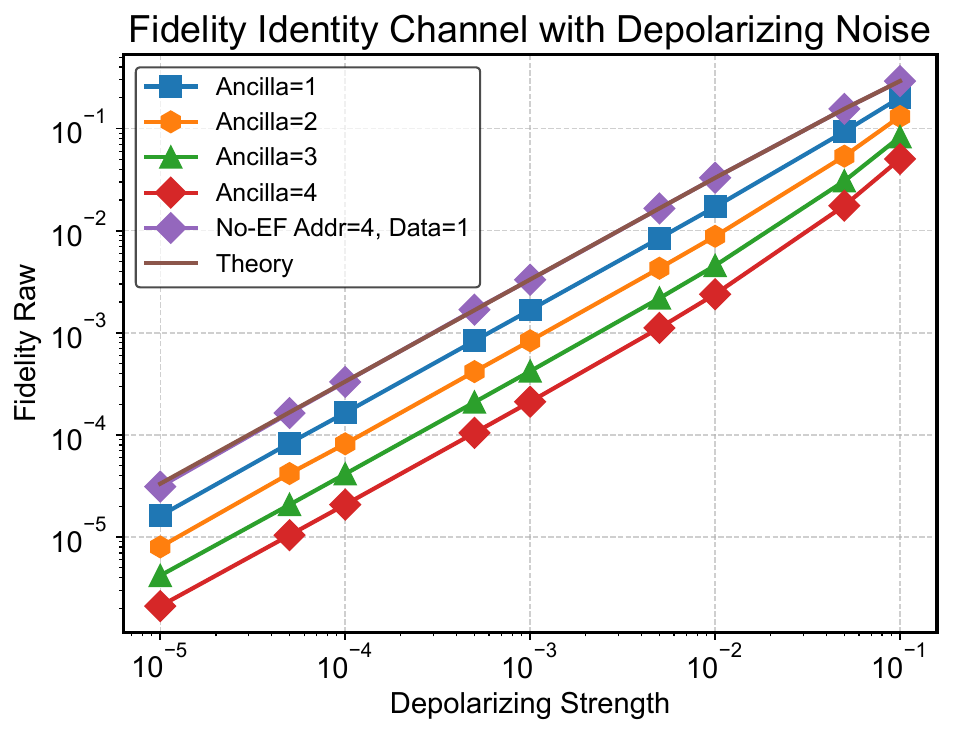}
\caption{Error filtration with increasing ancilla register size, corresponding to higher EF levels. Results confirm that higher EF levels consistently improve fidelity, in agreement with theoretical predictions.}
\label{fig:Same system size under depolarizing noise}
\end{figure}

We further tested the effect of increasing the error filtration level by adding more ancilla qubits. As shown in Fig.~\ref{fig:Same system size under depolarizing noise}, higher EF levels systematically suppress infidelity, and the simulation results align with theoretical predictions.

Taken together, these tests validate both the noise integration and EF protocol within the sparse simulator. Depolarizing noise reproduces the correct analytical decay, EF suppresses infidelity in the expected regime, and multi-register configurations behave consistently with theoretical predictions. Moreover, we observe that EF performance begins to fail once the raw infidelity approaches $\sim 0.1$, the same breakdown point later observed in BB-QRAM simulations. This confirms that the sparse simulator is both reliable in simple models and consistent with the behavior of the full QRAM simulator, paving the way for its integration in large-scale studies.

\section{Breakdown of Ideal Error Filtration Scaling in QRAM Simulations \label{Sec: Derivation of EF SI}}
To establish the baseline for error filtration (EF) suppression, we begin by defining infidelity. Let the ideal quantum operation be a unitary $U$ acting on the pure input state $\ket{\psi}$, resulting in $\ket{U\psi} \equiv U\ket{\psi} = \ket{\Psi}$. In the presence of noise, this operation is replaced by a noisy channel $\mathcal{U}$, which acts on the input density matrix $\rho = \ket{\psi}\bra{\psi}$ as
\begin{equation}
\mathcal{U}(\rho) = \mathcal{U}(\ket{\psi}\bra{\psi}).
\end{equation}
The fidelity of the noisy output relative to the ideal state is then
\begin{equation}
F_0 = \bra{\Psi} \mathcal{U}(\rho) \ket{\Psi},
\end{equation}
and the corresponding infidelity is defined as $(1 - F)_0 = 1 - F_0$. This serves as the reference quantity against which EF suppression will be measured.

Under ideal EF conditions, the infidelity is expected to be suppressed by a factor of $2^T$, where $T$ denotes the EF level, corresponding to the use of $T$ ancilla qubits and $2^T$ repetitions of the noisy operation. However, in practice, this ideal ratio is not always realized. A key theoretical question is therefore under what conditions the suppression ratio approaches $2^T$, and what prefactors or corrections limit its effectiveness. The following analysis begins with the explicit state evolution for the simplest case $T=1$.

\subsection{Case $T=1$: Single-Level EF Evolution}

We derive the output state and resulting fidelity for the EF protocol with filtration depth $T=1$, following the full circuit evolution step by step.

\paragraph{State preparation.}
The initial state consists of a control qubit in $\ket{0}$, a memory register $\rho_\psi$, and an ancilla register $\rho_\phi$:
\begin{equation}
\rho_0 = \ket{0}\bra{0} \otimes \rho_\psi \otimes \rho_\phi.
\end{equation}
Applying a Hadamard gate to the control yields
\begin{equation}
H \ket{0} = \frac{1}{\sqrt{2}} (\ket{0} + \ket{1}), 
\quad 
H \ket{1} = \frac{1}{\sqrt{2}} (\ket{0} - \ket{1}),
\end{equation}
so the full system becomes
\begin{equation}
\rho_1 = \frac{1}{2} (\ket{0}\bra{0} + \ket{0}\bra{1} + \ket{1}\bra{0} + \ket{1}\bra{1}) \otimes \rho_\psi \otimes \rho_\phi.
\end{equation}

\paragraph{First 0-controlled-SWAP.}
Applying a controlled-SWAP between memory and ancilla gives
\begin{align}
\rho_2 = \tfrac{1}{2} \Big[ &\ket{0}\bra{0} \otimes \rho_\phi \otimes \rho_\psi
+ \ket{0}\bra{1} \otimes \mathrm{SWAP}(\rho_\psi \otimes \rho_\phi) \nonumber \\
&+ \ket{1}\bra{0} \otimes (\rho_\psi \otimes \rho_\phi)\,\mathrm{SWAP}
+ \ket{1}\bra{1} \otimes \rho_\psi \otimes \rho_\phi \Big].
\end{align}

\paragraph{Noisy operation.}
After applying the noisy channel $\mathcal{U}$:
\begin{align}
\rho_3 = \tfrac{1}{2} \Big[ &\ket{0}\bra{0} \otimes \rho_\phi \otimes \mathcal{U}(\rho_\psi)
+ \ket{0}\bra{1} \otimes \mathcal{U}(\mathrm{SWAP}(\rho_\psi \otimes \rho_\phi)) \nonumber \\
&+ \ket{1}\bra{0} \otimes \mathcal{U}((\rho_\psi \otimes \rho_\phi)\mathrm{SWAP})
+ \ket{1}\bra{1} \otimes \rho_\psi \otimes \mathcal{U}(\rho_\phi) \Big].
\end{align}

\paragraph{Second 0-controlled-SWAP.}
Undoing the previous swap gives
\begin{align}
\rho_4 = \tfrac{1}{2} \Big[ &\ket{0}\bra{0} \otimes \mathcal{U}(\rho_\psi) \otimes \rho_\phi
+ \ket{0}\bra{1} \otimes \mathrm{SWAP}\cdot \mathcal{U}(\mathrm{SWAP}(\rho_\psi \otimes \rho_\phi)) \nonumber \\
&+ \ket{1}\bra{0} \otimes \mathcal{U}((\rho_\psi \otimes \rho_\phi)\,\mathrm{SWAP})\cdot \mathrm{SWAP}
+ \ket{1}\bra{1} \otimes \rho_\psi \otimes \mathcal{U}(\rho_\phi) \Big].
\end{align}

\paragraph{First 1-controlled-SWAP.}
\begin{align}
\rho_5 = \tfrac{1}{2} \Big[ &\ket{0}\bra{0} \otimes \mathcal{U}(\rho_\psi) \otimes \rho_\phi
+ \ket{0}\bra{1} \otimes \mathrm{SWAP}\cdot \mathcal{U}(\mathrm{SWAP}(\rho_\psi \otimes \rho_\phi))\cdot \mathrm{SWAP} \nonumber \\
&+ \ket{1}\bra{0} \otimes \mathrm{SWAP}\cdot \mathcal{U}((\rho_\psi \otimes \rho_\phi)\,\mathrm{SWAP})\cdot \mathrm{SWAP}
+ \ket{1}\bra{1} \otimes \mathcal{U}(\rho_\phi) \otimes \rho_\psi \Big].
\end{align}

\paragraph{Noisy operation.}
Applying $\mathcal{U}$ again:
\begin{align}
\rho_6 = \tfrac{1}{2} \Big[ &\ket{0}\bra{0} \otimes \mathcal{U}(\rho_\psi) \otimes \mathcal{U}(\rho_\phi)
+ \ket{0}\bra{1} \otimes \mathcal{U}(\mathrm{SWAP}\cdot \mathcal{U}(\mathrm{SWAP}(\rho_\psi \otimes \rho_\phi))\cdot \mathrm{SWAP}) \nonumber \\
&+ \ket{1}\bra{0} \otimes \mathcal{U}(\mathrm{SWAP}\cdot \mathcal{U}((\rho_\psi \otimes \rho_\phi)\,\mathrm{SWAP})\cdot \mathrm{SWAP})
+ \ket{1}\bra{1} \otimes \mathcal{U}(\rho_\phi) \otimes \mathcal{U}(\rho_\psi) \Big].
\end{align}

\paragraph{Second 1-controlled-SWAP.}
Undoing the swap yields
\begin{align}
\rho_7 = \tfrac{1}{2} \Big[ &\ket{0}\bra{0} \otimes \mathcal{U}(\rho_\psi) \otimes \mathcal{U}(\rho_\phi) \nonumber + \ket{0}\bra{1} \otimes \overbrace{\mathcal{U}(\mathrm{SWAP}\cdot \mathcal{U}(\mathrm{SWAP}(\rho_\psi \otimes \rho_\phi))\cdot \mathrm{SWAP})\cdot \mathrm{SWAP}}^{\text{term}_{01}} \nonumber \\
&+ \ket{1}\bra{0} \otimes \underbrace{\mathrm{SWAP}\cdot \mathcal{U}(\mathrm{SWAP}\cdot \mathcal{U}((\rho_\psi \otimes \rho_\phi)\,\mathrm{SWAP})\cdot \mathrm{SWAP})}_{\text{term}_{10}} \nonumber + \ket{1}\bra{1} \otimes \mathcal{U}(\rho_\psi) \otimes \mathcal{U}(\rho_\phi) \Big].
\end{align}
By expanding $\mathcal{U}(\cdot) = \sum_i K_i (\cdot) K_i^\dagger$ and rearranging, we obtain
\begin{equation}
\text{term}_{01} = \text{term}_{10} = \sum_{i,j} K_j \rho_\psi K_i^\dagger \otimes K_i \rho_\phi K_j^\dagger.
\end{equation}

\paragraph{Final Hadamard and post-selection.}
Applying a final Hadamard and post-selecting on the $\ket{0}$ outcome yields the success probability
\begin{equation}
P_S^{(1)} = \frac{1}{2}\left(1 + \operatorname{Tr}[\text{term}_{01}] \right),
\end{equation}
with the interference term
\begin{equation}
\operatorname{Tr}[\text{term}_{01}] = \sum_{i,j} \bra{\psi} K_j^\dagger K_i \ket{\psi}\, \mathrm{Tr}(\rho_\phi K_i^\dagger K_j).
\end{equation}

After tracing out the ancilla, the post-selected memory state is
\begin{equation}
\rho_\psi' = \frac{1}{2P_S^{(1)}}\left( \mathcal{U}(\rho_\psi) + \sum_{i,j} K_i \rho_\psi K_j^\dagger \cdot \mathrm{Tr}(\rho_\phi K_j K_i^\dagger) \right).
\end{equation}

\paragraph{Final fidelity.}
The fidelity after one EF round is therefore
\begin{equation}
F_1 = \frac{\bra{\Psi}\rho_1\ket{\Psi}}{P_S^{(1)}},
\end{equation}
with numerator
\begin{equation}
\bra{\Psi}\rho_1\ket{\Psi} = \frac{1}{2}F_0 + \frac{1}{2}\sum_{i,j} \bra{\psi} U^\dagger K_i \ket{\psi}\bra{\psi}K_j^\dagger U\ket{\psi}\,\mathrm{Tr}(\rho_\phi K_i^\dagger K_j),
\end{equation}
and denominator
\begin{equation}
P_S^{(1)} = \frac{1}{2} + \frac{1}{2}\sum_{i,j} \bra{\psi} K_j^\dagger K_i \ket{\psi}\,\mathrm{Tr}(\rho_\phi K_i^\dagger K_j).
\end{equation}

\paragraph{Noise model approximation.}
To align with the original EF formalism, we adopt a noise model with Kraus operators
\begin{equation}
K_0 = \sqrt{1-\varepsilon}\,U, 
\qquad 
K_1 = \sqrt{\varepsilon}\,V, 
\qquad V \neq U,
\end{equation}
where $V$ is an arbitrary erroneous unitary. Expanding up to $\mathcal{O}(\varepsilon^2)$, the two sums over $i,j$ coincide, giving
\begin{equation}
\bra{\Psi}\rho_1\ket{\Psi} = \frac{1}{2}\left(F_0 + 2P_S^{(1)} - 1\right).
\end{equation}
Thus,
\begin{equation}
\frac{\bra{\Psi}\rho_1\ket{\Psi}}{P_S^{(1)}} 
= 1 - \frac{1}{2}\frac{1-F_0}{P_S^{(1)}}, 
\qquad 
(1-F)_1 = \frac{1}{2}\frac{1-F_0}{P_S^{(1)}}.
\end{equation}

\paragraph{Suppression ratio.}
The ratio between pre- and post-filtration infidelity is therefore
\begin{equation}
\frac{1-F_0}{1-F_1} = 2P_S^{(1)} = \left(1 + (1-\varepsilon)^2 + \mathcal{O}(\varepsilon^2)\right) \sim 2(1 - \varepsilon) + \mathcal{O}(\varepsilon^2).
\label{eqn: EF ratio explain SI}
\end{equation}
This shows that the ideal suppression factor of $2^T$ is only achievable when $P_S^{(T)} \approx 1$. 
In practice, accumulated noise in realistic QRAM circuits lowers the post-selection probability, thereby degrading EF performance.

\subsection{General $T$ \label{Subsec: General T Analysis}}
We now generalize the EF derivation to general $T$.  
Let $\rho^{(T)}_{\text{pre-PS}}$ denote the joint state of the control, memory, and ancilla registers prior to post-selection.  
Projecting onto the all-zero ancilla state
\begin{equation}
\Pi_0 = (\ket{0}\bra{0})^{\otimes T} \otimes I \otimes I,
\end{equation}
and renormalizing yields the post-selected joint state
\begin{equation}
\rho_{\psi,\phi} =
\frac{\Pi_0 \, \rho^{(T)}_{\text{pre-PS}} \, \Pi_0}
{\operatorname{Tr}\!\left(\rho^{(T)}_{\text{pre-PS}} \, \Pi_0\right)},
\end{equation}
where the denominator
\begin{equation}
P^{(T)}_S = \operatorname{Tr}\!\left(\rho^{(T)}_{\text{pre-PS}} \, \Pi_0\right)
\end{equation}
is the post-selection success probability. The reduced memory state is then given by $\operatorname{Tr}_{\phi}(\rho_{\psi,\phi})$.

\paragraph{Kraus expansion.}
For later convenience, we define the \emph{partial Kraus string}
\begin{equation}
\overline{K}_{i_t} = K_{i_T} K_{i_{T-1}} \cdots K_{i_{t+1}} K_{i_{t-1}} \cdots K_{i_1}.
\end{equation}
In this notation, the success probability can be written as
\begin{equation}
P^{(T)}_S  
= \frac{1}{2^T} 
+ \frac{1}{4^{T}} \sum_{i} \sum_{t=1}^{2^T} \sum_{\substack{q=1 \\ q \ne t}}^{2^T}  
\left( \bra{\psi} K_{i_q}^\dagger K_{i_t} \ket{\psi} \;
\mathrm{Tr}\!\left[ \rho_\phi \, \overline{K_{i_q}^\dagger} \, \overline{K_{i_t}} \right] \right),
\end{equation}
while the unnormalized post-selected memory state is
\begin{equation}
\tilde{\rho}^{(T)}  
= \frac{1}{2^T} \, \mathcal{U}\!\left( |\psi\rangle \langle \psi| \right)  
+ \frac{1}{4^T} \sum_{i} \sum_{t=1}^{2^T} \sum_{\substack{q=1 \\ q \ne t}}^{2^T}  
\left( K_{i_t} |\psi\rangle \langle \psi| K_{i_q}^\dagger \;
\mathrm{Tr}\!\left[ \rho_\phi \, \overline{K_{i_q}^\dagger} \, \overline{K_{i_t}} \right] \right).
\end{equation}

\paragraph{Favorable conditions.}
Under the simplifying assumptions introduced in Ref.~\cite{Gideon_EFiltration_2023}, the success probability reduces to
\begin{equation}
P^{(T)}_S  = \frac{1}{2^T} 
+ \frac{2^T - 1}{2^T} \sum_{i, j}  
\bra{\psi} K_{i}^\dagger K_{j}  \ket{\psi} \;
\bra{\phi} K_{j}^\dagger K_{i}  \ket{\phi}.
\end{equation}
If the noise channel includes a Kraus operator of the form
\begin{equation}
K_0 = \sqrt{1 - \varepsilon}\, U,
\end{equation}
with $U$ the ideal target unitary, then the diagonal contribution ($i=j=0$) gives
\begin{equation}
\bra{\psi} K_{0}^\dagger K_{0}  \ket{\psi} \;
\bra{\phi} K_{0}^\dagger K_{0}  \ket{\phi} = (1 - \varepsilon)^2.
\end{equation}
The remaining terms contribute non-negatively if we assume the memory register and ancilla register share the same initial state, i.e., $\ket{\psi} = \ket{\phi}$,
\begin{equation}
\bra{\psi} K_{i}^\dagger K_{j}  \ket{\psi} \;
\bra{\psi} K_{j}^\dagger K_{i}  \ket{\psi} 
= \big|\bra{\psi} K_{i}^\dagger K_{j} \ket{\psi}\big|^2 \geq 0.
\end{equation}

\paragraph{Lower bound. \label{para: low_bound}}
It follows that
\begin{equation}
P^{(T)}_S \;\geq\; \frac{1}{2^T} + \frac{2^T - 1}{2^T}(1 - \varepsilon)^2 
\;\geq\; 1 - 2\varepsilon\Bigl(1 - \frac{1}{2^T}\Bigr) 
\;\geq\; 1 - 2\varepsilon.
\end{equation}
Thus, in some cases the EF success probability can be bounded below by a constant independent of $T$.  
This is highly desirable, as it implies that infidelity can be suppressed in a near-deterministic manner.  
Moreover, as $T$ increases, the failure rate $1-P^{(T)}_S$ approaches $2\varepsilon$, showing that EF performance saturates at this constant level, making it particularly effective for QRAM error suppression at large $T$. Both the dynamic bound $2\varepsilon\left(1 - \frac{1}{2^T}\right)$ and the constant bound $2\varepsilon$ are demonstrated in the Fig.~\ref{fig:Ratio_Scaling_and_Bound}.

\paragraph{Post-selected fidelity.}
After EF of level $T$, the normalized memory state is
\begin{equation}
\rho^{(T)} = \frac{\tilde{\rho}^{(T)}}{P^{(T)}_S},
\end{equation}
and the infidelity is
\begin{equation}
(1 - F)_{T} = 1 - \bra{U\psi} \rho^{(T)} \ket{U\psi}
= \frac{P_S^{(T)} - \bra{U\psi} \tilde{\rho}^{(T)} \ket{U\psi}}{P_S^{(T)}}.
\end{equation}
Since the numerator evaluates to $\tfrac{1}{2^T}(1-F_0) + \mathcal{O}(\varepsilon^2)$, we obtain the key scaling relation
\begin{equation}
\frac{1-F_0}{1-F_T} = 2^T P_S^{(T)}.
\end{equation}
This recovers the ideal $2^T$ suppression factor only in the limit $P_S^{(T)} \to 1$, and highlights the central role of the success probability in determining EF efficiency at finite noise levels.

\end{document}